\documentclass[prb,superscriptaddress, amsmath,amssymb,aps,twocolumn]{revtex4-2}

\usepackage{graphics,epsf}

\usepackage{graphicx}
\usepackage{dcolumn}
\usepackage{bm}
\usepackage[colorlinks=true, allcolors=blue]{hyperref}
\usepackage{cleveref}

\usepackage{amsmath}
\usepackage{amssymb}
\usepackage{amsfonts}

\usepackage{epstopdf}
\usepackage[T1]{fontenc}
\usepackage[latin9]{inputenc}
\usepackage{amsbsy}
\usepackage[T1]{fontenc}
\usepackage[latin9]{inputenc}

\usepackage{bbm}
\usepackage{braket}
\usepackage{xcolor}
\usepackage{float}
\usepackage[normalem]{ulem}

\usepackage{hyperref}

\usepackage{natbib}
\usepackage{graphicx}
\usepackage{color}
\usepackage{gensymb}
\usepackage{float}
\usepackage{amsmath}
\usepackage{tabularx,graphicx}
\usepackage{epstopdf}
\usepackage{latexsym}
\usepackage{color, colortbl}
\usepackage{psfrag}
\usepackage{bbm}
\usepackage{bm}
\usepackage{blindtext}
\usepackage{dsfont}
\usepackage{feynmp}
\usepackage{slashed}
\usepackage{multirow}
\usepackage{appendix}
\usepackage{ragged2e}
\usepackage{bbold}
\usepackage{braket,mleftright}

\def \ba{\begin{align*}}
\def \ea{\end{align*}}

\newcounter{indice}

\renewcommand{\approx}{\simeq}

\usepackage{graphics,epsf}

\usepackage{graphicx}
\usepackage{dcolumn}
\usepackage{bm}
\usepackage[colorlinks=true, allcolors=blue]{hyperref}
\usepackage{cleveref}

\usepackage{amsmath}
\usepackage{amssymb}
\usepackage{amsfonts}

\usepackage{epstopdf}
\usepackage[T1]{fontenc}
\usepackage[latin9]{inputenc}
\usepackage{amsbsy}
\usepackage[T1]{fontenc}
\usepackage[latin9]{inputenc}

\usepackage{bbm}
\usepackage{braket}
\usepackage{xcolor}
\usepackage{float}
\usepackage[normalem]{ulem}

\usepackage{hyperref}

\usepackage{natbib}
\usepackage{graphicx}
\usepackage{color}
\usepackage{gensymb}
\usepackage{float}
\usepackage{tabularx,graphicx}
\usepackage{epstopdf}
\usepackage{latexsym}
\usepackage{color, colortbl}
\usepackage{psfrag}
\usepackage{bbm}
\usepackage{bm}
\usepackage{blindtext}
\usepackage{dsfont}
\usepackage{feynmp}
\usepackage{slashed}
\usepackage{multirow}
\usepackage{appendix}
\usepackage{ragged2e}
\usepackage{bbold}
\usepackage{braket,mleftright}

\def \be{\begin{equation}}
\def \ee{\end{equation}}
\def \ba{\begin{array}}
\def \ea{\end{array}}
\def \bea{\begin{eqnarray}}
\def \eea{\end{eqnarray}}

\def \ba{\begin{align*}}
\def \ea{\end{align*}}

\renewcommand{\approx}{\simeq}

\begin{document}

\title{Mixed Triplet-Singlet Order Parameter in Decoupled Superconducting 1H Monolayers of Transition-Metal Dichalcogenides}

\author{Avior Almoalem}
\thanks{These authors contributed equally}
\affiliation{Department of Physics and Materials Research Laboratory, Grainger College of Engineering, University of Illinois at Urbana-Champaign, Urbana, IL, USA}

\author{Sajilesh Kunhiparambath}
\thanks{These authors contributed equally}
\affiliation{Department of Physics, Technion, Haifa, 3200003, Israel}

\author{Roni Anna Gofman}
\affiliation{Department of Physics, Technion, Haifa, 3200003, Israel}
\author{Yuval Nitzav}
\affiliation{Department of Physics, Technion, Haifa, 3200003, Israel}
\author{Ilay Mangel}
\affiliation{Department of Physics, Technion, Haifa, 3200003, Israel}
 
\author{Nitzan Ragoler}
\affiliation{Department of Physics, Technion, Haifa, 3200003, Israel}
\author{Jun Fujii}
\affiliation{Istituto Officina dei Materiali (IOM)-CNR, Area Science Park, S.S.14, Km 163.5, 34149 Trieste, Italy}
\author{Ivana Vobornik}
\affiliation{Istituto Officina dei Materiali (IOM)-CNR, Area Science Park, S.S.14, Km 163.5, 34149 Trieste, Italy}

\author{Francois Bertran}
\affiliation{SOLEIL Synchrotron, \v{L}Orme des Merisiers, D\'epartementale 128, 91190 Saint-Aubin, France}

\author{Amit Kanigel}
\affiliation{Department of Physics, Technion, Haifa, 3200003, Israel}

\author{Jonathan Ruhman}
\affiliation{Department of Physics, Bar-Ilan University, 52900, Ramat Gan, Israel}

\author{Vidya Madhavan}
\affiliation{Department of Physics and Materials Research Laboratory, Grainger College of Engineering, University of Illinois at Urbana-Champaign, Urbana, IL, USA}


\begin{abstract}
Understanding the emergence of unconventional superconductivity, where the order parameter deviates from simple isotropic s-wave pairing, is a central puzzle in condensed matter physics. Transition-metal dichalcogenides (TMDCs), though generally regarded as conventional superconductors, display signatures of this unusual behavior and thus provide a particularly intriguing platform to explore how exotic states arise. Here we investigate the misfit compound (SnS)$_{1.15}$(TaS$_2$), a heterostructure composed of alternating SnS and 1H-TaS$_2$ layers. Using transport, photoemission, and scanning tunneling spectroscopy, we demonstrate that the SnS layers effectively decouple the TaS$_2$ into electronically isolated 1H sheets. In this limit, the tunneling density of states reveals a clear two-gap superconducting spectrum with T$_c \approx 3.1$ K. A theoretical model based on lack of inversion symmetry and finite-range attraction reproduces the observed multi-gap structure as a mixed singlet-triplet state. These results establish misfit compounds as a powerful platform for studying unconventional superconductivity in isolated 1H layers and for realizing multiple uncoupled superconductors within a single crystal.

\end{abstract}

\maketitle

\section{INTRODUCTION}

The origin of unconventional pairing states remains one of the most compelling open questions in superconductivity (SC). While most superconductors host simple, isotropic s-wave Cooper pairs, analogous to spherical atomic orbitals, many materials depart from this paradigm. In such systems, the superconducting order parameter can have reduced symmetry, sign changes, or nodes, enabling a rich variety of emergent phenomena. These include nodal quasiparticles with characteristic power-law thermodynamics, spin-triplet states~\cite{mackenzie2003superconductivity,maeno2001,pustogow2019}, spatially modulated pair-density waves~\cite{berg2007,li2007,tranquada2008}, and topological phases supporting Majorana excitations and chiral edge currents~\cite{nayak2008,kitaev2001,alicea2012}. Prominent examples are the d-wave cuprates~\cite{tsuei2000pairing,scalapino1995}, certain heavy-fermion compounds~\cite{stewart1984,ott1983}, layered oxide superconductors~\cite{maeno2001,ishida1998,takagi1988}, and more recently, van der Waals (vdW) heterostructures~\cite{cao2018,sharpe2019,Ribak2020,zhou2021superconductivity}. Despite extensive experimental evidence, the microscopic ingredients driving such pairing remain elusive. 

It is widely believed that strong, short-range Coulomb repulsion and the resulting magnetic fluctuations are crucial for these pairing states~\cite{scalapino1995,monthoux2007,anderson1987,si2010heavy}. This suggests a connection between non-s-wave superconductivity and phenomena like strong correlations and quantum criticality~\cite{varma2002,abanov2003quantum,coleman2005,lohneysen2007,schattner2016ising,chubukov2020interplay}. However, not all candidate materials fit neatly into these categories, and even when they do, the precise mechanisms driving the pairing remain elusive.

A compelling platform for investigating these unconventional states is found in the transition metal dichalcogenide (TMDC) family~\cite{nakamura2017odd,lane2022identifying,hamill2021two,xi2016ising}. These are van der Waals materials, essentially stacks of two-dimensional crystals, which can exist in various polymorphs. The 2H polymorphs, such as 2H-NbSe$_2$, 2H-TaS$_2$, and 2H-TaSe$_2$, are superconductors with a $T_C$ ranging between 0.7 and 7 K~\cite{navarro2016enhanced,DelaBarrera2018,hamill2021two,simon2024transition}. While they are often described by BCS theory, they also exhibit a low-temperature charge density wave (CDW) transition, or at least a significant softening of their phonon spectra towards such an order~\cite{sun2024twisted,kim2025anomalous}. The basic 1H layers possess a non-centrosymmetric structure and strong spin-orbit coupling, which results in strong spin-valley locking (SVL)~\cite{Zhang2014,Riley2014,DelaBarrera2018,Bawden2016,AlmoalemSpin}.

In stark contrast, the 1T polymorphs of the same compounds undergo a much stronger CDW transition, leading to significant band reconstruction. For example, 1T-TaS$_2$ becomes insulating at low temperatures and is a candidate quantum spin liquid~\cite{law20171t,ribak2017gapless,klanjvsek2017high}. These two behaviors converge in the 4Hb polymorph of TaS$_2$, a unique hybrid structure composed of an alternating stack of 1H-TaS$_2$ and 1T-TaS$_2$ layers~\cite{di1973preparation,Ribak2020}. This unique stacking is believed to be the origin of the host of unconventional phenomena~\cite{liu2024magnetization,watson2025folded}, including multi-component and topological superconductivity~\cite{Nayak2021,Nayak2023,persky2022magnetic,silber2024two,AlmoalemLP},  absent in its individual constituent layers.

Recent measurements on exfoliated 2H-TaS$_2$ flakes have further complicated this picture by showing evidence for a multi-gap superconducting state~\cite{simon2024transition}, while a monolayer grown by molecular beam epitaxy (MBE) has signs of an f-wave order parameter~\cite{Vano23}. This leads to a crucial overarching question: what drives the emergence of an unconventional order parameter in these materials, and is the exotic superconducting state intrinsic to the 1H layer, or is it strongly influenced by its environment?
 
To answer this question and differentiate between an isolated monolayer and a system involving strongly correlated layers, we performed a low-temperature study using bulk and local probes of the superconducting phase in (SnS)$_{1.15}$(TaS$_{2}$). Misfit compounds are heterostructures with a miss-match between the lattice vectors of the different layers, effectively decoupling them into true two dimensional systems~\cite{devarakonda2020clean}. Interestingly, the low energy density of states (DOS) in the superconducting phase of the 1H layers is characterized by a multi-gap structure. Our theoretical model predicts that this structure emerges from the elimination of the inter-layer hopping, which brings about an admixture of singlet-triplet superconducting states. Our results, derived from various techniques, suggest that the SnS layer acts as a buffer layer between the 1H layers, effectively turning the material into stacked 1H-TaS$_2$ monolayers. This conclusion is also evident from the sample T$_C$=3.1~K. We also observe the formation of superconductivity in the SnS layer, which seems to be intrinsic to the layer and not formed by proximity from the 1H layers.

\section{multi-gap spectrum in 1h monolayer }

\begin{figure*}[t]
\includegraphics[trim= 0cm 0cm 0cm 0cm,clip=true,width=1\textwidth]{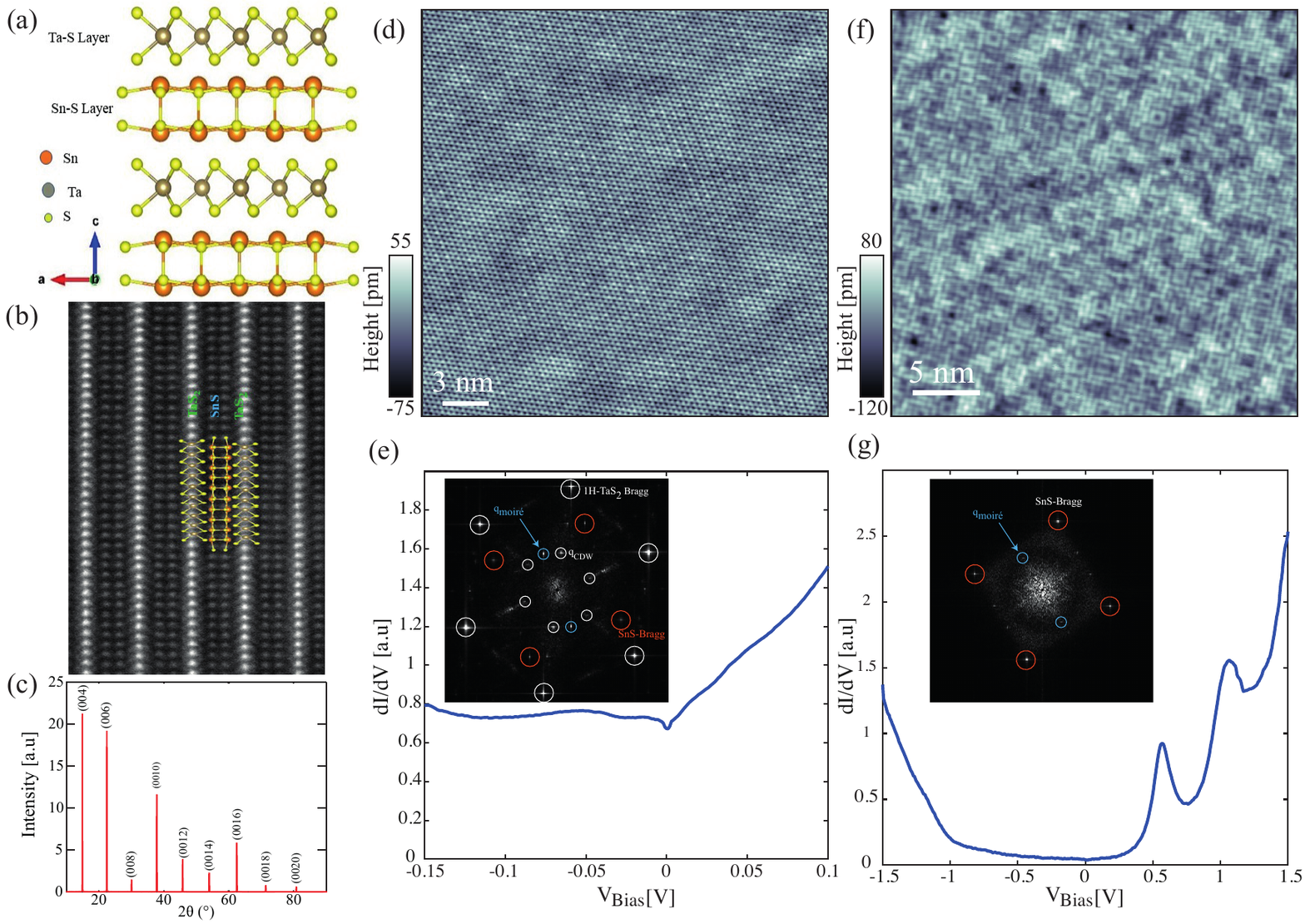}
\caption{(a) Crystal structure of (SnS)$_{1.15}$(TaS$_{2}$) viewed along the $b$ axis.  
(b) High-resolution TEM image showing the alternating stacking of TaS$_{2}$ and SnS layers.  
(c) Diffraction data from a (SnS)$_{1.15}$(TaS$_{2}$) single crystal.  
(d) Topography of 1H-TaS$_{2}$ layer in (SnS)$_{1.15}$(TaS$_{2}$) ($I_{\mathrm{set}} = 250$ pA and $V_{\mathrm{Bias}} = -30$ mV).  
(e) Spectra taken at the same location as (c). The spectrum is reminiscent of that of the 1H-TaS$_{2}$ layer in 4Hb-TaS$_{2}$. The small dip at the Fermi level is reminiscent of the SC gap. The modulation amplitude and step size smear the gap, leaving only a small dip. ($I_{\mathrm{set}} = 300$ pA, $V_{\mathrm{Bias}} = -150$ mV, $V_{\mathrm{mod}} = 0.5$ mV, $f = 870$ Hz). Inset: FFT of the topography shown in (d); 1H-TaS$_{2}$ Bragg and CDW peaks marked by white circles, SnS Bragg peaks in red circles, and the moir\'{e} peak marked with a cyan arrow. 
(f,g) Same for the SnS layer. The lack of a clear square lattice in the topography is probably due to damage to the layer in the cleaving process, where some atoms remain on the opposing cleaved surface. For (f) $I_{\mathrm{set}} = 250$ pA and $V_{\mathrm{Bias}} = -1.5$ V; for (g) $I_{\mathrm{set}} = 150$ pA, $V_{\mathrm{Bias}} = -1.5$ V, $V_{\mathrm{mod}} = 6$ mV, $f = 870$ Hz.
}
\label{Fig1}
\end{figure*}

The compound (SnS)$_{1.15}$TaS$_2$ consists of alternating layers of SnS and 1H layers, as shown in Fig.~\ref{Fig1}a. Bulk SnS is a semiconductor with an indirect band gap of 1.3 eV~\cite{SnS}, and 1H-TaS$_2$ is a metal that undergoes consecutive CDW and SC transitions. The SnS layers adopt a distorted NaCl structure with a double layer of tin and sulfur, while the 1H layers crystallize in a non-centrosymmetric structure, where Ta atoms are trigonal prismatically coordinated by six sulfur atoms. 

Unlike in 4Hb-TaS$_2$, where trigonal prismatic Ta-S layers are rotated by 180$^{\circ}$ degrees relative to each other, here the 1H layers maintain the same orientation, making the crystal lattice lack overall inversion symmetry. Furthermore, the 1H-TaS$_2$ layers are shifted by half a unit cell along the \(b\)-axis.
Due to the different symmetries of the two lattices, an irrational ratio exists between the lattice constants along one of the in-plane directions while maintaining a perfect stacking order along the c-axis, thus creating a misfit structure.

The samples' high quality is evident in the peak sharpness in X-ray diffraction measurements. The existence of the (00$l$) reflections confirms that the $c$-axis is perpendicular to the sample surface. High Resolution Transmission Electrons Microscopy (HRTEM) results confirm the crystal structure of alternate stacking of 1H and SnS layers, Fig.~\ref{Fig1}b,c. 

Upon cleaving, the exposed surface includes areas of both SnS and 1H-TaS$_2$ terminations. Figures~\ref{Fig1}(d,e) show the topography and high-energy tunneling spectra of the 1H termination, respectively. Bragg, CDW, and  moir\'{e} peaks are marked in the FFT shown in the inset of Fig.~\ref{Fig1}e. High energy spectra are similar to previously measured 1H layer in the heterostructures 4Hb-TaS$_2$~\cite{Nayak2021}. The CDW Q vector is similar to other studies with $Q_\mathrm{CDW}=0.3543\cdot Q_\mathrm{Bragg}$. Figures~\ref{Fig1}(f,g) similarly show the topography and spectra of the SnS layer. Bragg peaks of the square lattice, with the same moir\'{e} peak, are marked in the FFT. The moir\'{e} peaks in both layers originate from the mismatch and rotation of the two basic unit cells. The moir\'{e} peaks were also observed in a high-temperature STM study of the same material~\cite{Li2024}. 

In the SnS layer, $dI/dV$ high-energy spectrum is reminiscent of an insulator, as expected from the semiconducting SnS. The semiconducting gap, which exists from -0.5~eV up to 0.5~eV, is not full and has a finite DOS at the Fermi level. The residual DOS is distributed across the SnS termination, indicating that it is an intrinsic property of the layer.

\begin{figure}[!]
\includegraphics[width=1 \columnwidth]{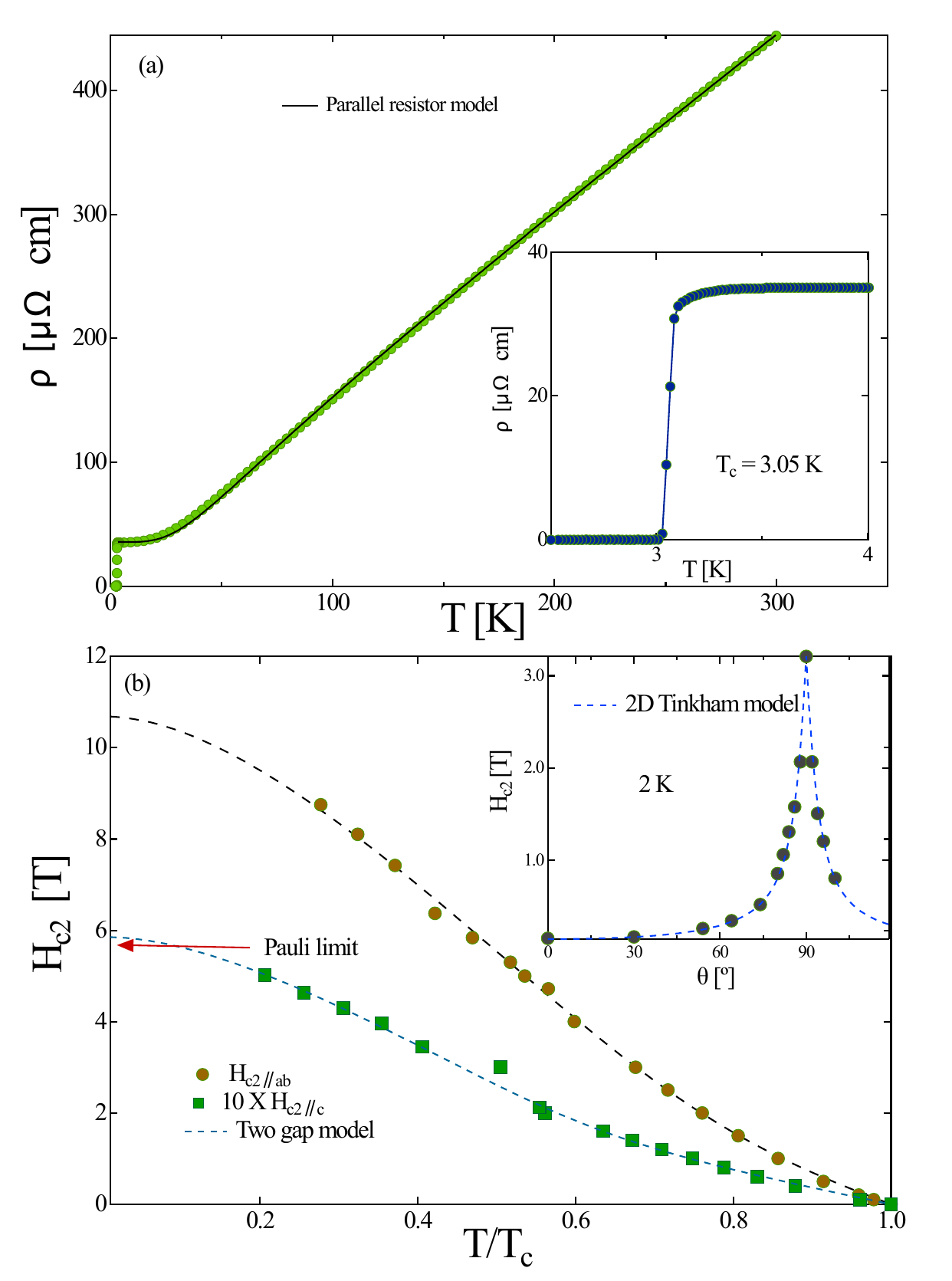}
 \caption{ (a) Electrical resistivity for the (SnS)$_{1.15}$(TaS$_{2}$) in the range 1.7 K $\geq T \geq$ 300 K. The dotted line shows the fit to the data using the parallel resistor model. The inset shows the expanded view, showing the drop in resistivity around 3.14 K. Temperature dependence of resistivity measured at different applied fields for out-of-plane. (b) Temperature dependence of upper-critical field for out-of-plane (green squares) and in-plane directions (brown circles). The dotted line shows the fitting using the two-gap model. The inset shows the angular dependence of the upper-critical field fitted using the 2D-Tinkham model~\cite{Tinkham}.}
\label{fig:STS_Res}
\end{figure}

Electrical transport measurements were performed on single crystals in a four-probe configuration, and the resistivity as a function of temperature is shown in Fig.~\ref{fig:STS_Res}a. The monotonic decrease in resistivity emphasizes the metallic nature of the material. Interestingly, no sign of the CDW transition appears in resistivity measurements, although CDW peaks are clearly visible in the FFT of the 1H topography. The high residual resistivity ratio (RRR),  $\rho_{300}/\rho_{10} = 12$, is comparable to that of other misfit compounds and serves as another indicator of the high quality of the crystals. At low temperatures, a sharp drop in resistivity occurs around $T$=3.05 K, indicating the SC transition (inset). This marks a considerable increase in $T_c$ compared to pristine 2H-TaS$_{2}$ ($T_{c}$=0.8 K), bringing it closer to the value observed for monolayer TaS$_{2}$ ($T_{c}$=3.4 K)~\cite{TaS2mono2}.

We define the critical field as the field at which resistivity drops to 50\% of its normal state value. Temperature dependence of both in- and out-of-plane critical fields is shown in Fig.~\ref{fig:STS_Res}b, with the raw data shown in Supplementary Figure S1. The inset shows the angular dependence of the critical field at 2K, with the data fitted to the 2D Tinkham model~ \cite{Tinkham}. We find an anisotropy ratio of 19.

Remarkably, we find an increasing concave temperature dependence of $H_\mathrm{c2}$ near $T_\mathrm{c}$ for both field orientations, instead of the conventional Werthamer-Helfand-Hohenberg (WHH) behavior. 
This behavior is most commonly attributed to the presence of two superconducting gaps, as observed in MgB$_{2}$~\cite{MgB2}, YNi$_{2}$B$_{2}$C~\cite{YNBC}, LuNi$_{2}$B$_{2}$C~\cite{YNBC2}, and 2H-NbSe$_{2}$~\cite{NbSe2}, although alternative explanations have also been proposed~\cite{hc21,hc22,hc23,hc24}.
A two-gap model, described in the supplementary material, is used to fit the data. The fit results are shown as dashed lines in Fig.~\ref{fig:STS_Res}c. We find that the in-plane critical field surpasses the Pauli limit by a factor of 1.95, an indicator of Ising superconductivity~\cite{Ising_SC}.

\begin{figure}[!]
\includegraphics[width=1\columnwidth]{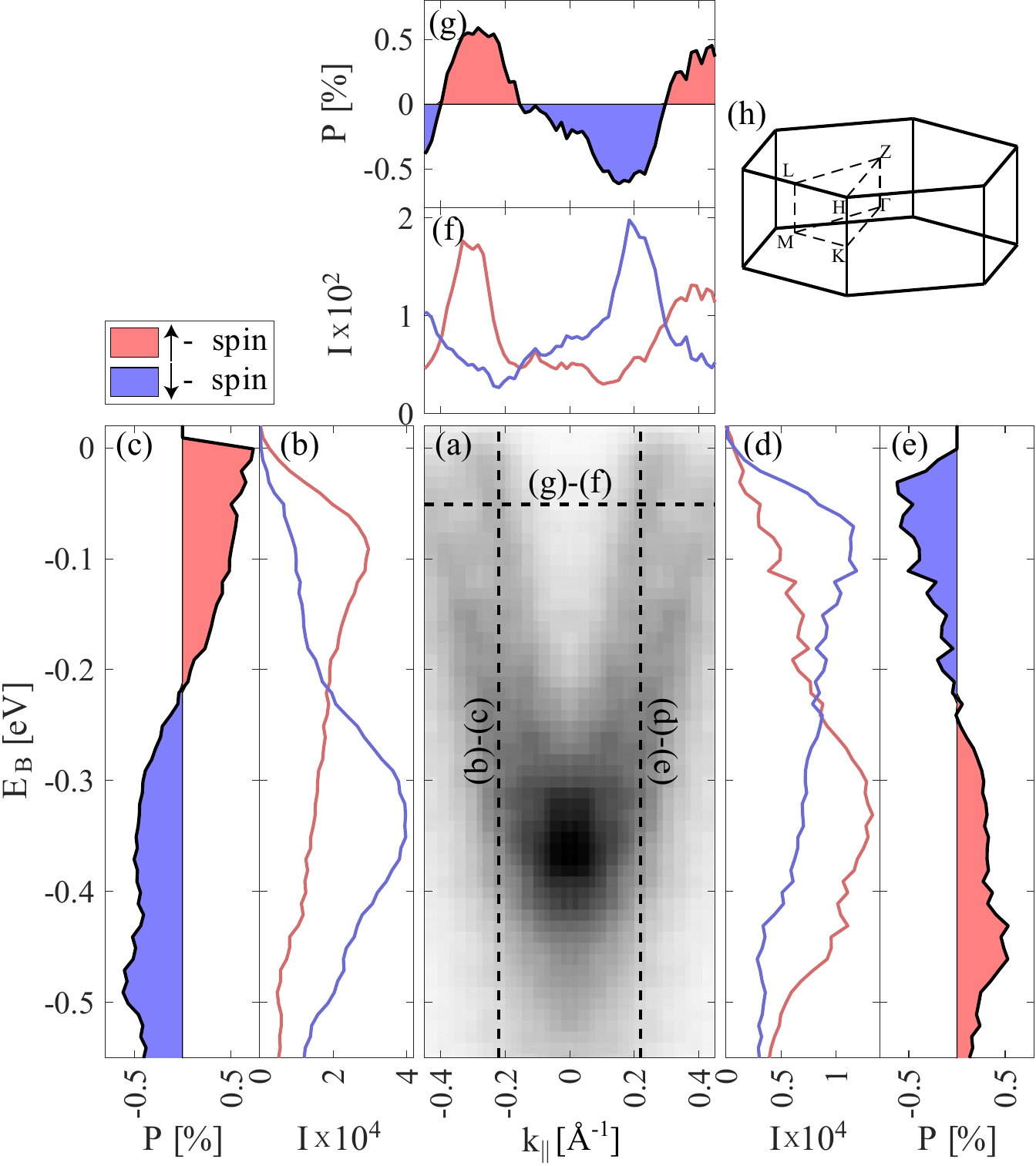}
\caption{(a) Symmetrized ARPES dispersion parallel to the in-plane high-symmetry line K-M-K', measured at 25 eV (corresponds to $k_z$=2.96 $\text{\AA}^{-1}$) and is approximately the midpoint along the M-L high-symmetry line. 
Dashed black lines mark the cuts used for spin-resolved measurements. The corresponding results are shown in panels (b-g). (b-c) Left vertical line: (b) Spin-resolved energy-distribution curves of spin-up (red) and spin-down (blue),  and (c) spin polarization vs energy (black), areas filled in red/blue indicate positive/negative polarization. (d-e) Same as b-c for the right vertical line. (f-g) Horizontal line: (f) Spin-resolved momentum-distribution curves of spin-up (red) and spin-down (blue),  and (g) spin polarization (black) vs momentum. (h) Hexagonal Brillouin zone of the 1H-TaS$_2$ with high-symmetry points labeled.}
\label{fig:SpinPol}
\end{figure}

Ising SC is a result of Spin-valley locking, a phenomenon where the electron's spin and valley degrees of freedom are intrinsically coupled due to strong spin-orbit interaction and broken inversion symmetry. This coupling ensures that the spin orientation is uniquely tied to the valley index, resulting in opposite spin polarizations at different valleys (e.g.,  K and K' points in the Brillouin zone). Cooper pairs forming the Ising SC inherit the spin-valley locking and can withstand large in-plane magnetic fields.
  
We show spin valley locking in (SnS)$_{1.15}$(TaS$_{2}$) directly, by using spin-resolved angle resolved photoemission spectroscopy (ARPES) measurements, shown in Fig.~\ref{fig:SpinPol}. Spin-polarized data measure the projection of the spin in the out-of-plane direction for each state in momentum space. The Fig.~\ref{fig:SpinPol}a shows ARPES spectra along the in-plane projected K-M-K' momentum direction (see Brillouin zone in Fig.~\ref{fig:SpinPol}h), with clear polarization of the bands seen in both the energy distribution curves (EDCs) (panels b-e) and the momentum distribution curve (MDC). MDC is taken 50meV below the Fermi level (panels f-g), thus giving the polarization for the occupied portion of the band structure close to the Fermi level. This is consistent with previous measured TMDCs with a 2H crystal structure~\cite{AlmoalemSpin,Bawden2016,Riley2014,Zhang2014}.

\begin{figure}[!]
\includegraphics[trim= 0cm 0cm 0cm 0cm,clip=true,width=1\linewidth]{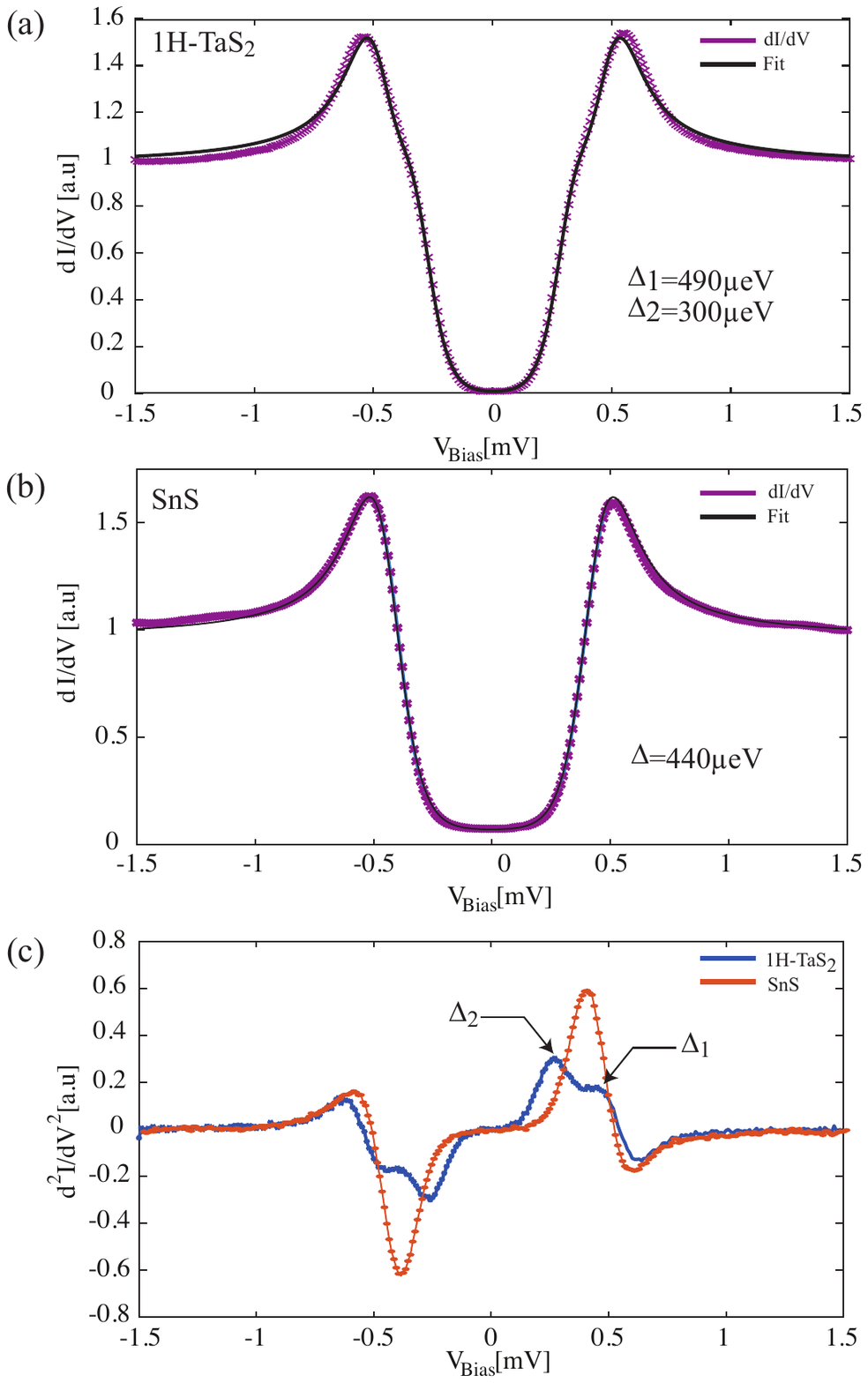}
\caption{ (a) The superconducting gap in 1H-TaS$_2$ layer. dI/dV spectrum obtained on the 1H-TaS$_2$ layer shown in Fig.~\ref{Fig1}c. The superconducting gap has no residual DOS, unlike the case of 4Hb-TaS$_2$. Overlying the data is a fit to a two gap function with different sizes. The clear change in the slope of the spectra around 300 $\mu$eV emphasize the two-gap behavior (I$_{set}$ = 50 pA and V$_{Bias}$ = -1.5meV, T=0.3K). (b) The superconducting gap in SnS layer. dI/dV spectrum obtained on the SnS layer shown in Fig.~\ref{Fig1}e. The SC gap is well fitted using a single BCS gap and a residual DOS (I$_{set}$ = 50 pA and V$_{Bias}$ = -2meV, T=0.3K). (c) $d^2I/dV^2$ taken from the data in a and b. The two arrows mark the energies of the two gaps acquired on 1H-TaS$_2$ layer, matching the energies of the change in the slope of the dI/dV.}
\label{Figgap}
\end{figure}

We now turn to low temperature, low energy STM/S data. We directly measure the superconducting gap at low temperatures with high resolution. The spectra obtained from the 1H layer reveal a distinct gap structure that resembles a two-gap spectrum of different magnitudes. Fitting the data yields an excellent match with a two-gap model (see Supplementary for details), as shown in Fig.~\ref{Figgap}a, whereas a single-gap model provides a poor fit (see Supplementary Figure S2). The spectrum remains uniform through the scan area (see Supplementary Figure S3), with the average spectrum displayed in Fig.~\ref{Figgap}a. Pronounced coherence peaks are observed at 490 $\mu$eV, above and bellow the Fermi level. The slope of the spectra, representing the change in DOS and thus the disappearance of states in the gap, steepens around 300 $\mu$eV therefore signaling the onset of the smaller gap. To highlight this feature, we present the first derivative of the spectra, which reveals two distinct "kinks" corresponding to the onset of the two gaps, as shown in Fig.~\ref{Figgap}c.

Alongside the 1H layer we perform similar measurements in the SnS layer. Bulk SnS is semiconducting and therefore not suppose to host superconductivity. Nevertheless, a clear gap can be observed in the spectrum acquired in the SnS layer as shown in Fig.~\ref{Figgap}b. The magnitude of the gap in the SnS layer is almost the same as the larger gap in the 1H layer. To confirm that the gap on the SnS layer is of superconducting origin, we measure the critical field in both layers (Supplementary Figure S4) and image vortices in the SnS layer (Supplementary Figure S5). This contrasts 4Hb-TaS$_2$, where the non-SC 1T layer shows no vortices, although a vortex exists below the layer as measured on a step edge in an STM study~\cite{Nayak2021}. 
 
 Imaging the vortex allows us to extract the coherence length by fitting the Fermi-level DOS to $N(E_F,x) = N_0 e^{-x/\xi} + N_\infty$~\cite{Nayak2021}, yielding $\xi_{\text{vortex}} \approx 35 \pm 5$ nm, where x is the distance from the core and $N_0$ and $N_\infty$ denote the DOS at the vortex core and far from the vortex, respectively. This value is consistent with the coherence length estimated from the upper critical field, $H_{c2} = \Phi_0 / (2\pi \xi^2)$, which gives $\xi_{H_{c2}} \approx 30$ nm at $T = 400$ mK.

The finite DOS at the Fermi level in SnS is intrinsic to the layer, and is similar to what is found in the 1H terminations of 4Hb-TaS$_2$ ~\cite{Nayak2021}. Interestingly, in 4Hb-TaS$_2$, no gap is observed on the 1T terminations, despite the superconducting nature of the underlying 1H layer. This observation suggests that the SnS gap is not of proximity origin.

The transport measurements of the bulk, and the tunneling spectrum in the 1H layer are consistent with a multi-gap structure, and cannot be explained within a single-gap framework. Considering that ARPES results suggest the lack of inversion symmetry and a two-dimensional electronic structure, a theoretical model, grounded in simple symmetry considerations and realistic band structure parameters, of a multi-gap structure naturally follows.

\section{theory of spin triplet-singlet mixing}

Comparison of our tunneling density of states in the superconducting state with 2H-TaS$_2$ flakes~\cite{simon2024transition}, where sub-gap features were observed only in the monolayer limit, suggests a correlation between inter-layer coupling and the prominence of multi-gap features. Specifically, multi-gap behavior appears in SnS-TaS$_2$, monolayer 1H-TaS$_2$, and 4Hb-TaS$_2$~\cite{wang2024evidencemultibandgaplesssuperconductivity}--systems in which inter-layer coupling is effectively quenched, for example, via increased inter-layer separation. In contrast, the multi-gap spectral signature disappears in 2H-TaS$_2$, highlighting the potential role of inter-layer coupling. The inter-layer coupling non-negligible effect in bulk 2H-TaS$_2$, as oppose to other TaS$_2$ based heterostructures, was shown previously by spin-ARPES measurements~\cite{AlmoalemSpin}. This raises two key questions: Why does this material support multi-gap superconductivity? And why is it suppressed by inter-layer coupling?

\begin{figure}[!]
\includegraphics[width=1\columnwidth]{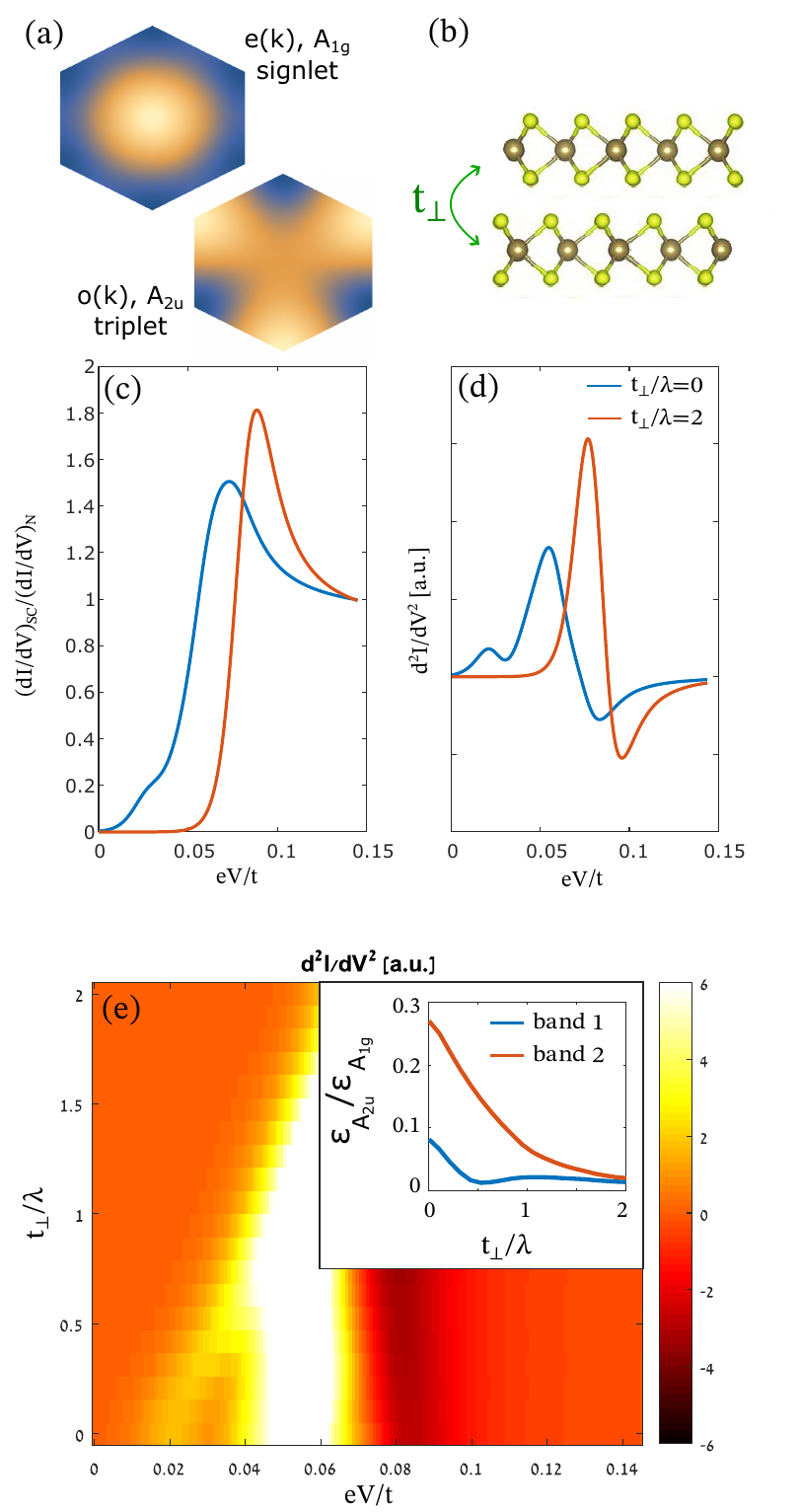}
\caption{(a) Schematic of the theoretical model. We consider 2H-TaS$_2$ with a varying interlayer hopping $t_\perp$ and Ising spin-orbit coupling $\lambda$. The attractive interactions in each layer are assumed to be finite range (on-site, NN and NNN). (b) Cartoon of our model with two 1H layers. the interlayer coupling is given by $t_\perp$. (c) The resulting tunneling DOS for $t_\perp=0$ (blue) and $t_\perp/\lambda = 2$ as a function of $eV/t$ in units of the hopping parameter (see supplement). (d) The  derivative  of the DOS shown in panel (b) showing two peaks corresponding to two gaps. (e) A color map of the derivative of the DOS for various values of the ratio $t_\perp/\lambda$. As can be seen the small gap continuously grows and eventually merges with the main gap around $t_\perp = 0.8 \lambda$. The inset shows the ratio between the triplet and singlet component in the gap functions $\varepsilon_{A_{2u}}/\varepsilon_{A_{1g}}$(see text) for the two bands of 2H-TaS$_2$ vs. $t_\perp/\lambda$, indicating that the mixing with the f-wave order parameter is the culprit behind the two-gap feature.}\label{fig:theory}
\end{figure}

A natural scenario that accounts for these observations is a singlet-triplet mixing due to local-inversion symmetry breaking~\cite{goryo2012possible}. To explore this idea, consider two one-dimensional irreducible representations (irreps) of the $D_3$ point group, fully symmetric at first: $A_{1g}$, which includes 
the constant function $1$, and the even function $e(\mathbf{k})$, and $A_{2u}$, which includes the odd function $o(\mathbf{k})$ (see Fig.~\ref{fig:theory}(a) and Ref.~\cite{goryo2012possible} for more details). Both are symmetric under $C_3$ rotations and mirror symmetry about the $\Gamma$-K line ($M_x$), but differ in their parity: $1$ and $e(\mathbf{k})$ are even under inversion, while $o(\mathbf{k})$ is odd.

The SC order parameter in a two-dimensional trigonal crystal can be decomposed into contributions from these irreps via:
\begin{equation}\label{eq:gap_function}\hat \Delta^{\sigma \sigma'}_\Gamma (\boldsymbol k) =  \sum_{\Gamma} \varepsilon_{\Gamma}\;f_\Gamma(\boldsymbol k)\; \chi_{\sigma \sigma'}^\Gamma  \,, \end{equation}
where $f_\Gamma(\mathbf{k})$ denotes the orbital form factor, $\chi^\Gamma$ is the corresponding spin matrix, and $\varepsilon_{\Gamma}$ are the amplitudes. For example, the even-parity ($s$-wave) basis function $f_{A_{1g}}(\mathbf{k})$ (a superposition of $1$ and $e(\mathbf{k})$) comes with the singlet matrix $\chi^{A_{1g}} = -i \sigma^y$, while the odd-parity ($f$-wave) function $f_{A_{2u}} = o(\mathbf{k})$ is paired with the triplet matrix $\chi^{A_{2u}} = -i \sigma^y \sigma^z$~\footnote{We do not consider triplet states with finite projection on the $z$-axis of the spin-angular momentum.}. 

In a 1H monolayer, the inversion symmetry is absent, the only symmetry that distinguishes $A_{1g}$ and $A_{2u}$ channels. As a consequence, these two channels mix. This is akin to the Stark effect, where an electric field along $z$ mixes the atomic $s$ and $p_z$ orbitals. Consequently, the eigenfunctions of the Hamiltonian include two superpositions of the above orbitals, which are split in energy. The same happens in 1H-TaS$_2$, the order parameter generically contains both singlet and triplet components. The relative strength of the components in the superposition, $\varepsilon_{A_{1g}}$ and $\varepsilon_{A_{2u}}$, depends on material parameters like spin-orbit coupling and pairing interactions. 

Coupling this monolayer to a second, inverted layer to form the 2H structure restores global inversion symmetry, although each monolayer still breaks inversion locally. In the limit of weak inter-layer coupling, it is reasonable to describe the order parameter as a superposition of monolayer states with singlet-triplet mixing~\footnote{The superposition of the two is obviously an eigen-function of inversion.}. In this way, one can construct gap functions that are eigenmodes of inversion but locally on each layer still exhibit singlet-triplet mixing.  
As inter-layer coupling increases however, the system becomes increasingly sensitive to global inversion symmetry, suppressing local inversion-breaking effects and consequently reducing their admixture.

To complete the picture, it is essential to establish a connection between singlet-triplet mixing and the emergence of multiple gaps in the quasi-particle spectrum. A pure $A_{1g}$ or $A_{2u}$ gap function is not expected to lead to such features and especially not to depend on inter-layer coupling. In contrast, the superposition of singlet and triplet leads to an imbalance of the  gap strength on $K$ and $K'$ points. This is because the superposition of odd and even functions adds up in some regions but cancel out in the others.  Therefore, when the weights $\varepsilon_{A_{1g}}$ and $\varepsilon_{A_{2u}}$ are comparable, the difference between the gaps can be significant, and a multi-gap feature in the density of states is expected. 

Our theoretical analysis shows that on-site Coulomb repulsion can significantly enhance the triplet component and by so enhance the weight $\varepsilon_{A_{2u}}$, which is crucial to reach the state where the variation of the gap is a prominent effect in the spectrum. 

The singlet-triplet admixture therefore provides a natural and symmetry-allowed mechanism for generating the multi-gap behavior observed in experiment and offers a compelling explanation for the empirical correlation between multi-gap superconductivity and inter-layer coupling in materials with locally broken inversion symmetry, such as 4Hb-TaS$_2$ or (SnS)$_{1.15}$(TaS$_{2}$).

To test this hypothesis, we construct a minimal, single-orbital tight-binding model~\cite{yu2025quantum} for the band structure of a single 2H-TaS$_2$ layer. The model incorporates three crucial ingredients:
(i) a tunable interlayer coupling $t_\perp$ (see Fig.~\ref{fig:theory}(b));
(ii) an Ising spin-orbit coupling $\lambda$; and
(iii) a finite-range pairing interaction, including on-site, nearest-neighbor, and next-nearest-neighbor attraction.
We suppress the on-site attraction relative to the longer-range terms due to screened Coulomb repulsion. All three ingredients are essential to explain the transition from multigap to single-gap superconductivity. 

In Fig.~\ref{fig:theory}.(c) we plot the resulting tunneling DOS for $t_\perp=0$ (blue) and $t_\perp/\lambda = 2$ as a function of $eV/t$, the bias voltage in units of the hopping parameter (see supplement). A clear two-gap feature emerges in the absence of inter-layer coupling, which disappears for large $t_\perp/\lambda$. In panel (d) we plot the derivative of the DOS, shown in panel (c), showing two peaks corresponding to two gaps. In panel (e) we plot a color map of the derivative of the DOS for various values of the ratio $t_\perp/\lambda$. As can be seen, the small gap continuously grows and eventually merges with the main gap around $t_\perp = 0.8 \lambda$. The inset shows the ratio between triplet and singlet ($\varepsilon_{A_{2u}}/\varepsilon_{A_{1g}}$), vs. $t_\perp/\lambda$. The two traces, red and blue, correspond to the two spin-degenerate bands of a single 1H layer. Thus, the inset shows that the two-gap feature is correlated with a large singlet-triplet mixing. Thus, the theoretical model successfully reproduces the main experimental observation: the multigap feature emerges in the limit where interlayer tunneling is suppressed. This suggests that, in the isolated limit, the origin of the two-gap feature is strong singlet-triplet mixing.

\section{Discussion}

The misfit compound (SnS)$_{1.15}$(TaS$_2$) provides a natural realization of electronically decoupled 1H-TaS$_2$ layers. The absence of $k_z$ dispersion~\cite{AlmoalemSpin} and appearance of spin-valley locking~\cite{Zhang2014} in ARPES data, and strong transport anisotropy all confirm that inter-layer hybridization is strongly suppressed, leaving an electronic structure characteristic of isolated 1H layers. In this regime, tunneling spectroscopy reveals a two-gap density of states.

This behavior can be placed in context with other TMDC systems. In bulk 2H-TaS$_2$, STM measurements report a single gap of about 0.28 meV~\cite{Bulk_2H_STM_1}. In MBE-grown monolayer 1H-TaS$_2$, a single $f$-wave-like gap has been reported~\cite{Vano23}, although with a low $T_c$ around 1 K. In contrast, exfoliated monolayers with $T_c$ near 3.5 K display two distinct gaps of approximately 0.25 and 0.46 meV~\cite{simon2024transition}, in close agreement with our findings.

Related behavior is also observed in 4Hb-TaS$_2$, where the 1H layers are decoupled by intervening 1T layers~\cite{AlmoalemSpin}. Specific heat~\cite{Ribak2020} and STM measurements~\cite{Nayak2021} show a residual density of states below the superconducting gap, consistent with a second gap that is destroyed by pair-breaking impurities~\cite{Ruhman_gapless,wang2024evidencemultibandgaplesssuperconductivity}. Thus, the observation of gapless subgap states in 4Hb-TaS$_2$ can be seen as another example of two-gap superconductivity. The overall picture emerging from these comparisons is that multi-gap superconductivity is an intrinsic property of isolated 1H-TaS$_2$ layers suppressed by inter layer coupling.

To investigate the origin of this behavior we constructed a theoretical model based on a tight-binding Hamiltonian of a 2H bilayer with varying inter-layer coupling, Ising spin-orbit coupling, finite-range attraction, and weak on-site repulsion, naturally accounts for this phenomenology. The resulting superconducting order is a triplet-singlet mixture of $f$-wave and $s$-wave components, which depends on the inter layer coupling. In the limit of strongly coupled layers, like the 2H system, the triplet component is small, and the gap is more or less uniform. In contrast, in the strongly coupled case, the $f$-wave component becomes comparable with the $s$-wave and leads to large variations in gap size across the Fermi pockets.

The absence of inversion symmetry in 1H layers is central to this result, and the framework should apply broadly to superconducting 1H-TMDCs such as NbSe$_2$. While bulk NbSe$_2$ shows two gaps due to an additional Se-derived band, recent studies on monolayer NbSe$_2$, and with S substitution support a role for inversion-symmetry breaking in generating multi component order parameter~\cite{mockli2018robust,steinberg2025transition,moreno2025gapless}. 
It is important to note that the model requires a substantial Coulomb repulsion such that the on-site attraction is reduced below the strength of the nearest- or next-nearest-neighbor attraction. This, in turn, implies that we are effectively considering a longer-range attractive interaction, which is plausible given the presence of soft phonon modes associated with the CDW transition.

While considering another explanation one may speculate that the two-gap structure arises from eliminating the inter-layer coupling and how it affects the band structure. However, as experiment shows~\cite{AlmoalemSpin} altering the inter-layer coupling produces only minor modifications to the band structure. For example, the $\Gamma$-point extremum is shifted by a few tens of meV over a bandwidth of 300 meV, while leaving the Fermi-level band structure essentially unchanged.

It is worth noting that our study indicates that the superconducting gap observed in the SnS layers is intrinsic and not induced by proximity to 1H-TaS$_2$. The nearly identical magnitude of the SnS gap to the larger 1H gap, the absence of a second proximity gap, and the high anisotropy of transport all argue against a proximity scenario. Moreover, theory predicts that a proximitized semiconductor-superconductor bilayer with Fermi surface mismatch should exhibit a strongly reduced induced gap~\cite{stanescu2022proximity}, unlike what we observe.

Interestingly, bulk SnS becomes superconducting under high pressure~\cite{Matsumoto2019} above 40 GPa, with $T_c$ up to 6 K, consistent with theoretical expectations that an enhanced density of states at the Fermi level can drive superconductivity. In the misfit compound, we propose that the structural arrangement effectively applies an internal pressure to the SnS layers, stabilizing an intrinsic superconducting phase even at ambient conditions.

In summary, (SnS)$_{1.15}$(TaS$_2$) emerges as a unique platform where decoupled 1H-TaS$_2$ layers exhibit multigap superconductivity and where SnS layers host an independent superconducting state. This duality not only makes misfit compounds valuable for studying 1H physics in isolation but also opens opportunities to explore coupled and uncoupled superconductors within a single crystalline system.

\section{Acknowledgments}
The work at UIUC was supported by a grant from the US Department of Energy, Office of Science, Basic Energy Sciences, under award number DE-SC0014335. V.M. acknowledges partial support from Gordon and Betty More Foundation's EPiQS Initiative through grant GBMF4860. The work at the Technion was supported by Israeli Science Foundation grant number ISF-1263/21. We acknowledge Elettra Sincrotrone Trieste for providing access to its synchrotron radiation facilities. This work has been partly performed in the framework of the nanoscience foundry and fine analysis (NFFA-MUR Italy Progetti Internazionali) facility. We acknowledge SOLEIL for provision of synchrotron radiation facilities. 

\newpage

\onecolumngrid

\section{Methods and materials}

\subsection{Material growth}
Single crystals of (SnS)$_{1.15}$(TaS$_{2}$) were grown using the chemical vapor transport (CVT) method. The starting materials, weighed in their nominal stoichiometric ratios, were thoroughly mixed, pre-sintered, and sealed in an evacuated quartz ampoule along with SnCl$_{2}$ as the transport agent. The sealed ampoule was then placed in a two-zone tube furnace with a temperature gradient of 100$^{\circ}$C, and the transport reaction was carried out for two weeks. Shiny, plate-like crystals formed in the cold end of the ampoule were collected and subsequently washed with ethanol to remove residual transport agent
\subsection{STM}

STM measurements were performed using a Unisoku STM at an instrument base temperature of 270mK, using chemically etched and annealed tungsten. Spectra were acquired using a standard lock-in technique at a frequency of 907 Hz.

\subsection{ARPES}
Spin-resolved ARPES data was acquired in the APE-LE beamline at Elettra, Italy. 
The light used at APE-LE is s polarized. Samples were cleaved in-situ and measured at temperatures ranging from 10 to 30 K. For the spin-resolved measurements, a VLEED detector was used. 

The polarization was calculated using a Sherman function of $S=0.27$, with the band polarization given by~\cite{Bawden2016,Riley2014,AlmoalemSpin}:
\begin{equation}
    P = \frac{1}{S} \frac{I_{\rm{pos}} - I_{\rm{neg}}}{I_{\rm{pos}} + I_{\rm{neg}}},
\end{equation}
where $I_{\rm{pos}}$ and $I_{\rm{neg}}$ are the measured intensities with the VLEED detector magnetized in the positive and negative directions, respectively.  
To average over temporal fluctuations in the photon flux and to remove systematic errors due to detector magnetization, we performed a measurement sequence with alternating field directions: positive, negative, negative, positive.  

Accordingly, the spin-resolved intensities are obtained as:
\begin{equation}
    I_\uparrow = \frac{1 + P}{2} \left( I_{\rm{pos}} + I_{\rm{neg}} \right) = \frac{I_{\rm{pos}}}{2} \left(1+\frac{1}{S}\right)+\frac{I_{\rm{neg}}}{2} \left(1-\frac{1}{S}\right),
\end{equation}
\begin{equation}
    I_\downarrow = \frac{1 - P}{2} \left( I_{\rm{pos}} + I_{\rm{neg}} \right)  = \frac{I_{\rm{pos}}}{2} \left(1-\frac{1}{S}\right)+\frac{I_{\rm{neg}}}{2} \left(1+\frac{1}{S}\right)
\end{equation}
where $I_\uparrow$ and $I_\downarrow$ denote the spin-up and spin-down components, respectively.  
Note that if the Sherman function were ideal ($S = 1$), then $I_\uparrow$ would equal $I_{\rm{pos}}$ and $I_\downarrow$ would equal $I_{\rm{neg}}$, as the detector would perfectly distinguish spin.

Photon-energy dependent ARPES data was measured at the CASSIOPEE beam-line at Soleil, France.
To determine the k$_z$ dispersion from photon-energy-dependent ARPES, we use the free electron final state approximation for normal emission~\cite{Riley2014,Almoalem2021link,AlmoalemSpin}:
\begin{equation}
    k_z = \sqrt{2m/\hbar^2 (E_{kin}+V_0)}
\end{equation}
where $\mathrm{V_0}=9 \; \rm{eV}$ is the inner potential~\cite{AlmoalemSpin} and $\mathrm{E_{kin}}$ is the kinetic energy of the photo-emitted electron.

\subsection{Transport}
Electrical transport measurements were carried out using a Quantum Design PPMS equipped with a dilution refrigerator, employing the standard four-probe technique. The sample was rotated relative to the applied magnetic field using an integrated rotator module for transport anisotropy measurements.

\bibliographystyle{apsrev4-2}
\bibliography{MainBib}

\end{document}


\preprint{APS/123-QED}

\title{Supplementary Material of Mixed Triplet-Singlet Order Parameter in Decoupled Superconducting 1H Monolayer of Transition-Metal Dichalcogenides}

\author{Avior Almoalem}
\thanks{These authors contributed equally}
\affiliation{Department of Physics and Materials Research Laboratory, Grainger College of Engineering, University of Illinois at Urbana-Champaign, Urbana, IL, USA}

\author{Sajilesh Kunhiparambath}
\thanks{These authors contributed equally}
\affiliation{Department of Physics, Technion, Haifa, 3200003, Israel}

\author{Roni Anna Gofman}
\affiliation{Department of Physics, Technion, Haifa, 3200003, Israel}
\author{Yuval Nitzav}
\affiliation{Department of Physics, Technion, Haifa, 3200003, Israel}
\author{Ilay Mangel}
\affiliation{Department of Physics, Technion, Haifa, 3200003, Israel}
 
\author{Nitzan Ragoler}
\affiliation{Department of Physics, Technion, Haifa, 3200003, Israel}
\author{Jun Fujii}
\affiliation{Istituto Officina dei Materiali (IOM)-CNR, Area Science Park, S.S.14, Km 163.5, 34149 Trieste, Italy}
\author{Ivana Vobornik}
\affiliation{Istituto Officina dei Materiali (IOM)-CNR, Area Science Park, S.S.14, Km 163.5, 34149 Trieste, Italy}

\author{Francois Bertran}
\affiliation{SOLEIL Synchrotron, \v{L}Orme des Merisiers, D\'epartementale 128, 91190 Saint-Aubin, France}

\author{Amit Kanigel}
\affiliation{Department of Physics, Technion, Haifa, 3200003, Israel}

\author{Jonathan Ruhman}
\affiliation{Department of Physics, Bar-Ilan University, 52900, Ramat Gan, Israel}

\author{Vidya Madhavan}
\affiliation{Department of Physics and Materials Research Laboratory, Grainger College of Engineering, University of Illinois at Urbana-Champaign, Urbana, IL, USA}


\date{\today}

\maketitle
\onecolumngrid
\renewcommand*{\figurename}{\textbf{Supplementary Figure} }

\setcounter{figure}{0} 
\section{Crystallography}
X-ray diffraction pattern of the single crystal is shown in  Fig.1. The (00$l$) reflections indicate the plane of the single crystal is perpendicular to the crystallographic c-axis and the sharpness of the peaks demonstrates the good quality of the single crystals.
This data was taken using Rigaku, SmartLab 9kW high resolution system, employing Cu K$_\alpha$ ($\lambda = 1.54056 \AA$) radiation. All of the reflection planes seen in the data are from the (00$l$) family, indicating a high quality single crystal with one crystallographic orientation. The peak narrow width also indicates a high purity sample.

\section{Resistivity}

Closely observing the resistivity data for 5 K $\leq$ T $\leq$ 300 K shows a saturating resistivity at high temperatures. Resistivity saturation is often observed in materials with multiple conduction channels, in particular for many intermetallic materials~\cite{Gunnarsson03}.  This characteristic behavior can be well described by a parallel resistivity model. At high temperatures, due to increased thermal energy, the mean free path of the electrons decreases and becomes close to a few interatomic spacings. The dominant scattering mechanism at these temperatures is due to electron-phonon interaction. The scattering cross section will no longer be linear in scattering perturbation. So the resistivity will not be proportional to mean square atomic displacement, which is proportional to temperature for harmonic potentials. Hence, with further increase in temperature, the slope of $\rho(T)$ is decreasing and eventually resistivity saturates. For a metallic material,  $\rho(T)$ can be written as $\rho_{i}(T) = \rho_{i,0} + \rho_{i,L}(T)$, where the first term indicates the residual resistivity that arises from impurities and disorder, which is temperature-independent. The second term addresses the temperature-dependent general resistivity which can be written according to the Bloch-Gruneison model as 

\begin{equation}
\rho_{i,L}(T) = C\left(\frac{T}{\Theta_{D}}\right)^{5}\int_{0}^{\Theta_{D}/T}\frac{x^{5}}{(e^{x}-1)(1-e^{-1})}dx
\label{eq:rhoiLT}
\end{equation}

where $\Theta_{D}$ is the Debye temperature and C is the material-dependent prefactor. Considering the resistivity saturation at high temperatures, the resistivity curve can be explained by

\begin{equation}
\rho(T) = \left[\frac{1}{\rho_{s}} + \frac{1}{\rho_{i}(T)} \right]^{-1}
\label{eq:rhoT}
\end{equation}

where $\rho_{s}$ is the saturation value of resistivity at high temperatures. The data show an excellent fit of this model as shown in Fig.2a, and yields Debye temperature $ \Theta_{D} $ =  168 $ \pm $ 2 K, C = 3.7 $ \pm $ 5 $   m\Omega $cm, residual resistivity $ \rho_{i,0} $ = 145 $ \pm $ 1 $ \mu\Omega $cm, $ \rho_{s} $ = 23302 $ \pm $ 4 $ \mu\Omega $cm.

\section{Critical field fitting}

To fit the critical fit data as in Fig.2c, we make use of the following parametric equation to extract the behavior, and extrapolated critical fields at T=0K.

\begin{equation}
ln\left(\frac{1}{t}\right) = \left[U(s) + U(\eta s) + \frac{\lambda_{0}}{w}\right] +   \left( \frac{1}{4} \left[U(s) - U(\eta s) - \frac{\lambda_{-}}{w}\right]^{2} +  \frac{\lambda _{12}\lambda _{21}}{w^{2}} \right) ^{1/2}
\end{equation} 

\begin{equation*}
H_{c2}   = \frac{ 2\phi_{0}Ts}{D_{e} } \;\;\;\;\;\;  \eta = \frac{D_{h}}{D_{e}}\;\;\;\;\;\; 
  U(s) = \psi(s+ 1/2) - \psi(1/2)\;\;\;\;\;\;\;\;\;\;\;\;\;\;
\end{equation*} 

\label{whh}

Here, $ \lambda_{-} $ = $ \lambda_{11} - \lambda_{22} $, $ \lambda_{0} = (\lambda_{-}^{2} + 4\lambda_{12}\lambda_{21})$, $ w = \lambda_{11}\lambda_{22} - \lambda_{21}\lambda_{12} $. The variables, $ \lambda_{11}, \lambda_{22}, \lambda_{21}, \lambda_{21} $ represents the matrix elements of the BCS coupling constants. $ D_{1} $ and $ D_{2} $ are the electron and hole diffusivity, respectively. $ \phi_{0} $ is the flux quantum and $ \psi (s) $ is the digamma function. The fitting of the experimental data using this equation, and extrapolating the upper critical field to T=0 K gives 0.59~T for out of plane magnetic field and 10.68~T for an in-plane field. The \cref{table} shows the fitting parameters obtained. The fitting has converged to $\lambda_{22}$ = $\lambda_{11}$, $\lambda_{12}$ = $\lambda_{21}$ with  $\lambda_{11}$ higher than $\lambda_{12}$, indicating the mostly decoupled nature of the bands. $T_c$ = 3.05 K was kept fixed during the fitting.   
\begin{table}[h!]
\centering
\begin{tabular}{ c  c  c  c  c  } 
\hline

type & $D_{1}$ & $\eta$ & $\lambda_{11}$ & $\lambda_{12}$ \\
\hline
\hline
$H_{c2 \parallel c}$ & 11.01 & 0.19 & 0.88 & 0.35\\
\hline
$H_{c2 \parallel ab}$ & 0.67 & 0.19 & 1.37 & 0.49\\
\hline
\end{tabular}
\caption{Fitting parameters from two-gap model for $H_{c2}$}
\label{table}
\end{table}

\begin{figure}[h!]
\includegraphics[width=1.0\columnwidth]{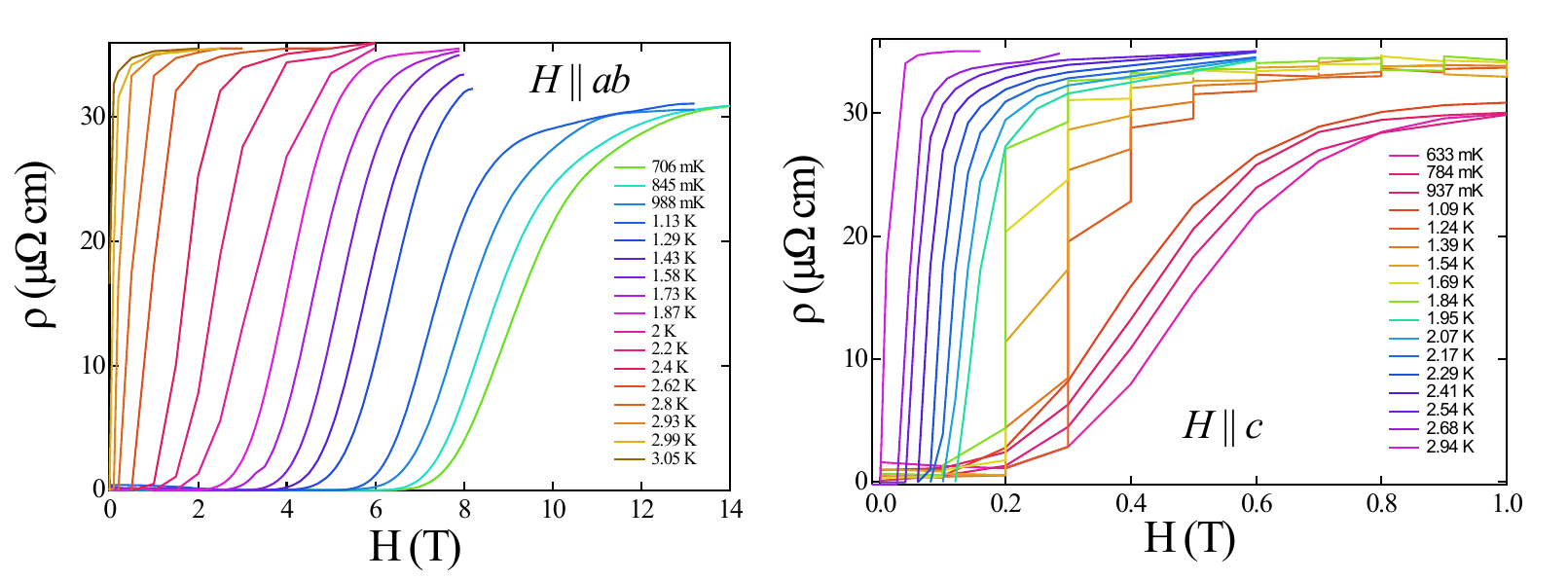}
\caption{ Resistivity vs field data taken at different temperatures for in-plane and out of plane direction }
\label{Fig_RH}
\end{figure}
\newpage
\section{Superconducting gap}

\subsection{Gap Fitting}

\begin{figure}[!h]
\includegraphics[trim= 0cm 0cm 0cm 0cm,clip=true,width=0.5\textwidth]{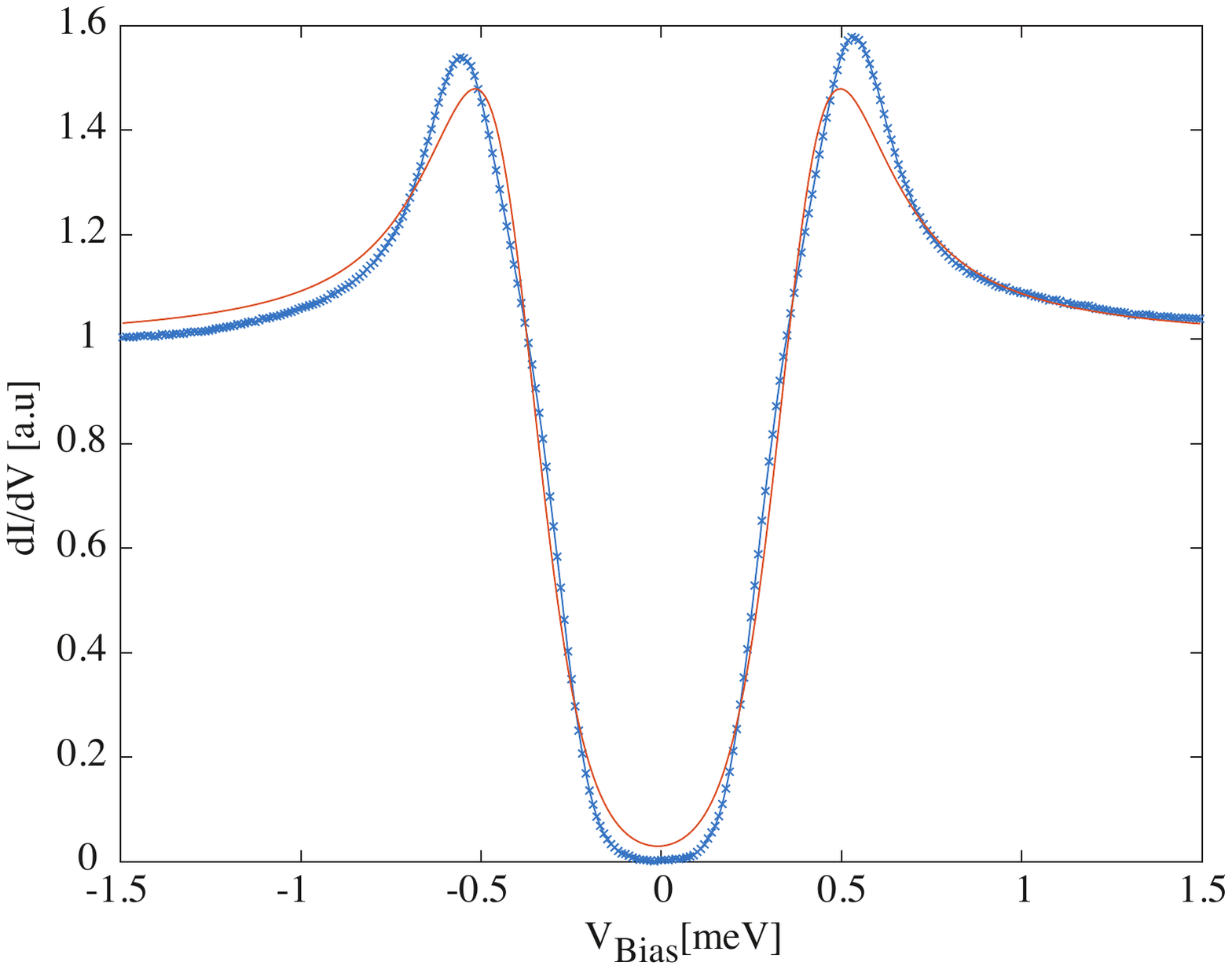}
\caption{Fitting the SC gap to a single BCS gap. Besides the residual denity of states arising from the considerably high effective temperature (900~mK) given by the fit, the slope of the DOS curve does not match the measured one, emphasizing the two-gap behavior measured on the 1H layer.}
\label{FigS4}
\end{figure}

We tried two simple models to fit the superconducting gap, the first was a single isotropic gap which yields poor results, Fig.\ref{FigS4}. For the single gap fit we use the function\cite{Vano23}:

\begin{align}\label{eq:fit1}
\frac{dI}{dV} (V) = \int_{-\infty}^{\infty} dE \rho_0 N_{0} \left|\mathfrak{R}\left[\frac{E}{\sqrt{E^2-\Delta^2}}\right]\right|\left( \frac{df(E+V, T_{eff})}{dV} \right)
\end{align}
where f is the Fermi-Dirac distribution, V is the bias voltage and $\Delta$ is the superconducting gap.

The two gap model which was used is given by:
\begin{equation}\label{eq:fit2}
\frac{dI}{dV} (V) = \int_{-\infty}^{\infty} dE \rho_0  \left(N_{1}\left|\mathfrak{R}\left[\frac{E}{\sqrt{E^2-\Delta_1^2}}\right]\right|+N_{2}\left|\mathfrak{R}\left[\frac{E}{\sqrt{E^2-\Delta_2^2}}\right]\right|\right)\left( \frac{df(E+V, T_{eff})}{dV} \right)
\end{equation}
where f is the Fermi-Dirac distribution, V is the bias voltage, $N_1$ and $N_2$ are different weights of the gaps, and $\Delta_{1,2}$ are the superconducting gaps.

\subsection{Gap Uniformity}
The superconducting gap in the sample is extremely uniform, indicating the clean sample, and the highe quality of the crystal. Moreover, it indicates the adatoms coming after the cleave process does not change the electronic properties, and do not interact with the superconducting phase.

\begin{figure}[!]
\includegraphics[trim= 0cm 0cm 0cm 0cm,clip=true,width=0.65\textwidth]{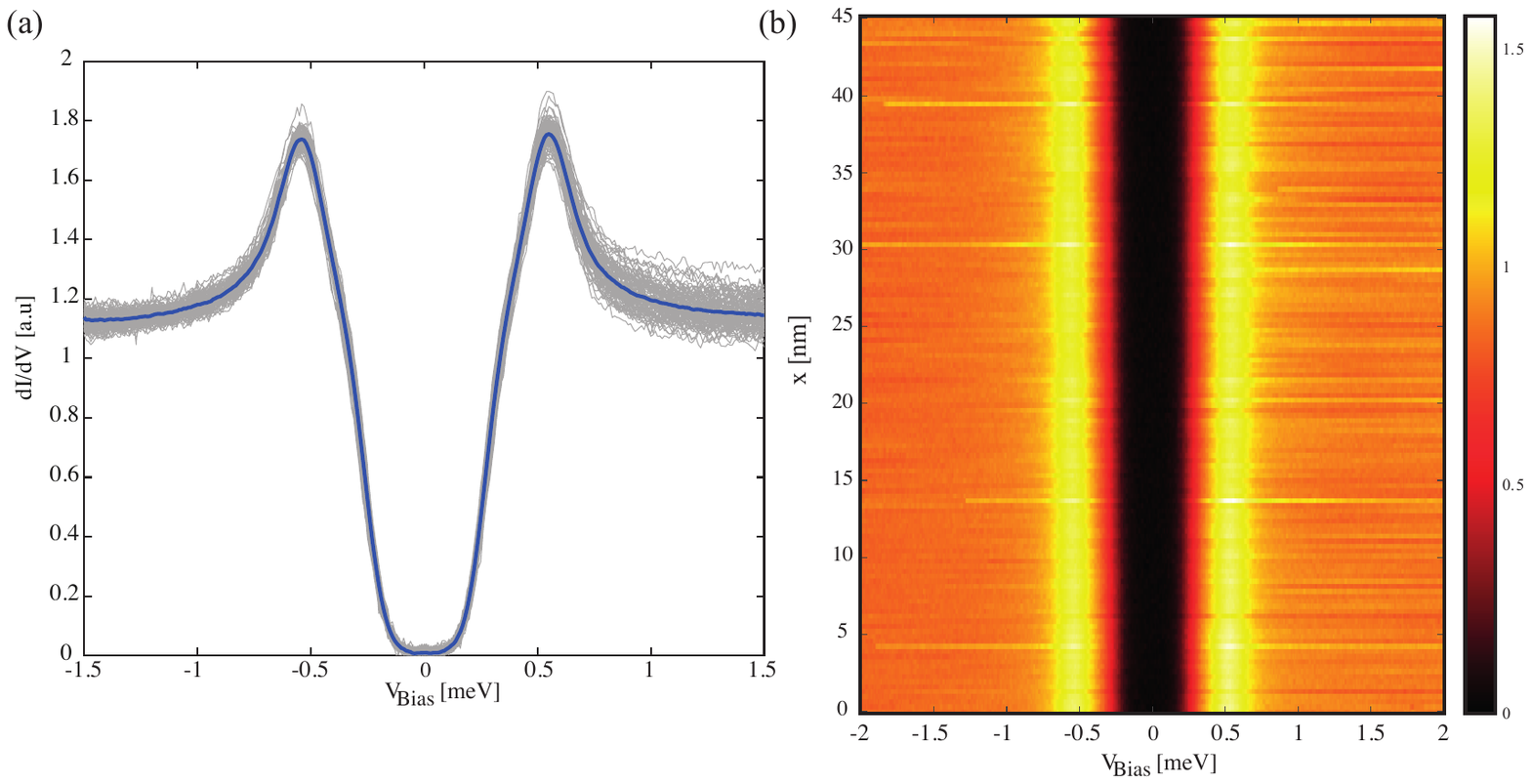}
\caption{(a) The superconducting gap in the 1H layer taken as an average of 150 different points. The blue curve is the average, and the light gray curves are the different spectra at different locations along a 45 nm line. I$_{set}$ = 50 pA and V$_{Bias}$ = -2meV, T=0.3K (b) False color plot showing the uniformity of the gap. The data is the same as in (a).}
\label{FigS3}
\end{figure}

\subsection{Magnetic field dependence of the gap}
\begin{figure}[!]
\includegraphics[trim= 0cm 0cm 0cm 0cm,clip=true,width=0.9\textwidth]{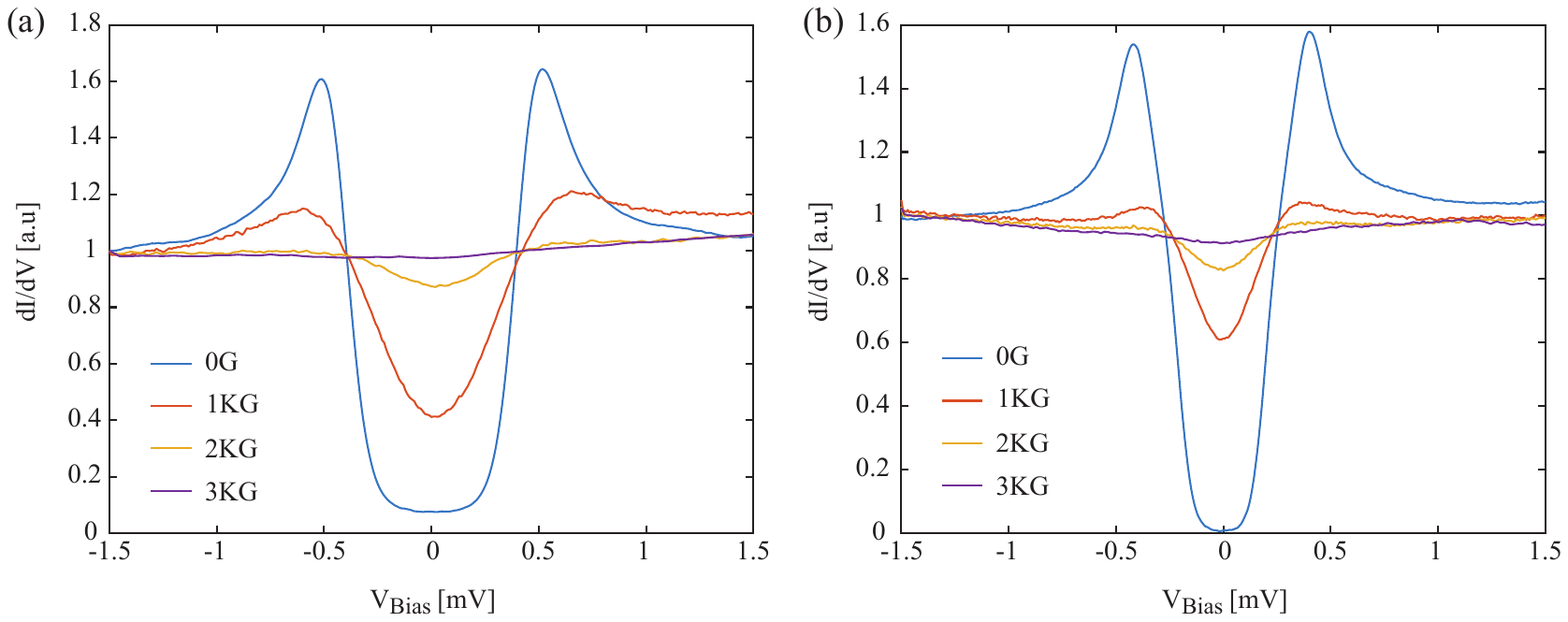}
\caption{Magnetic field dependence of the superconducting gap as measured in the SnS (a) and 1H-TaS$_2$ (b) layers, for B $\parallel$ c. The critical field measured is 3KG for both layers, and have excellent agreement with the critical field measured using transport at this effective temperature (T~0.5K). }
\label{Bdep}
\end{figure}

\begin{figure}[!h]
\includegraphics[trim= 0cm 0cm 0cm 0cm,clip=true,width=0.9\textwidth]{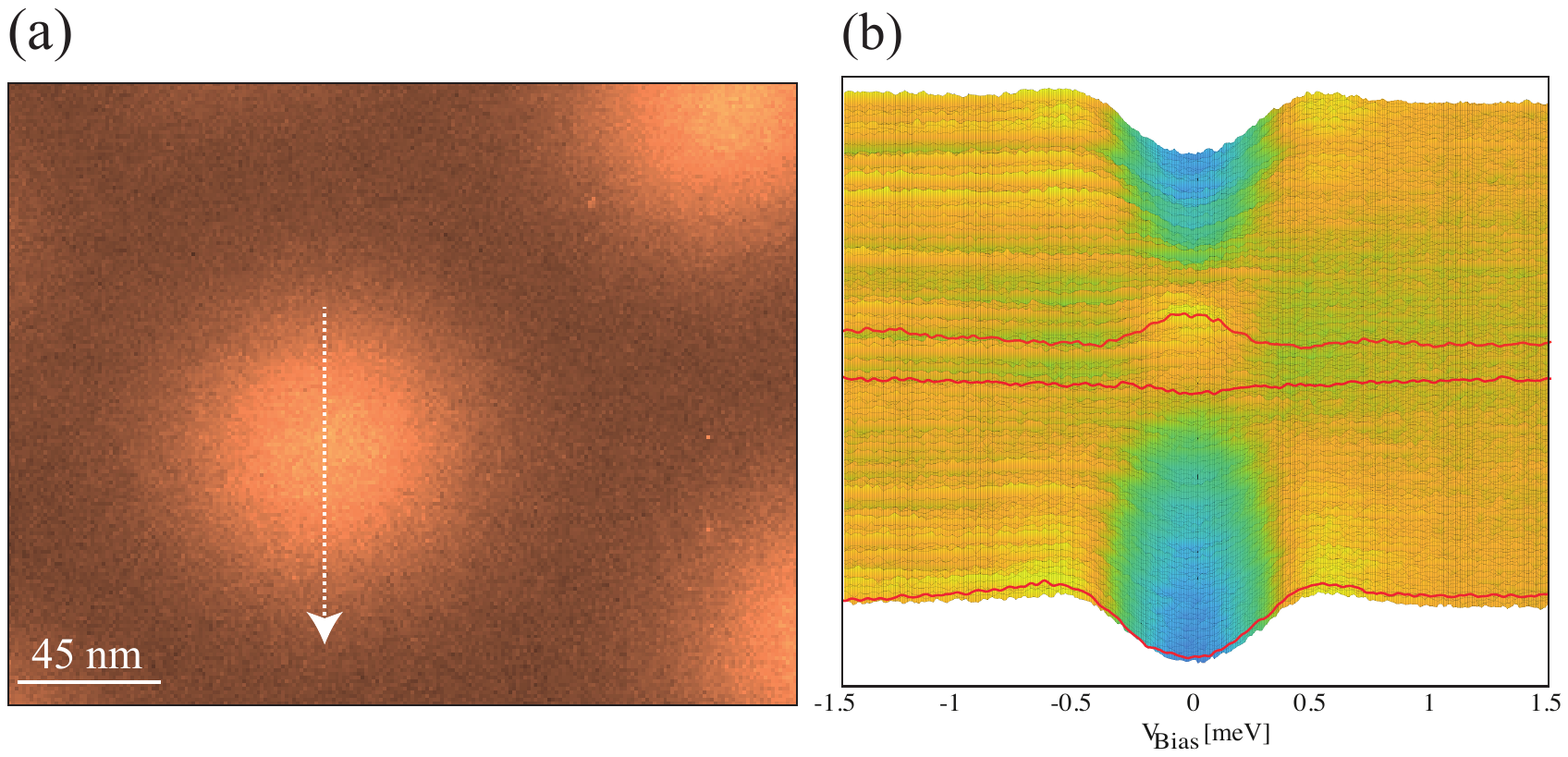}
\caption{(a) dI/dV map at E=0 acquired on a SnS layer at T=0.3K and B=0.1T (I$_{set}$=35pA, V$_{Bias}$ = -5mV). The vortex map shows that the gap in the SnS layer is of superconducting origin. Arrow mark the line cuts taken on the vortex at high resolution. (b) dI/dV spectra taken along line cut, and shows the enhanced density of states at the vortex core. }
\label{vortex}
\end{figure}

\newpage
\section{Vortex In the SnS layer}

we show the dI/dV map at E=0 in the SnS layer, at B=0.1T. The distance between the bright spots matches that expected from a B=0.1 T magnetic field (150 nm). The dI/dV spectra of the line cut as a function of position are seen in Fig.S\ref{vortex}b. A fit to the filling of the DOS in the gap has produced $\xi=45\pm10$nm, a coherence length that matches that extracted from H$_{c2}$ measurements. The dI/dV map at the Fermi level and other line cuts, not shown here, give an isotropic vortex shape. This evidence shows that the gap in the SnS layer is of superconducting origin and is intrinsic to the layer as opposed to a proximity gap due to the 1H layer. 

A clear filling of the gap is seen in the vortex core, with an increase in the density of states at its center. This extra dI/dV signal at the vortex core is probably due to Caroli-de Gennes-Matricon states, which is expected to exist in the core of every superconducting vortex.

\newpage
\section{Additional ARPES Data}

\begin{figure}[!h]
\includegraphics[width=0.3\columnwidth]{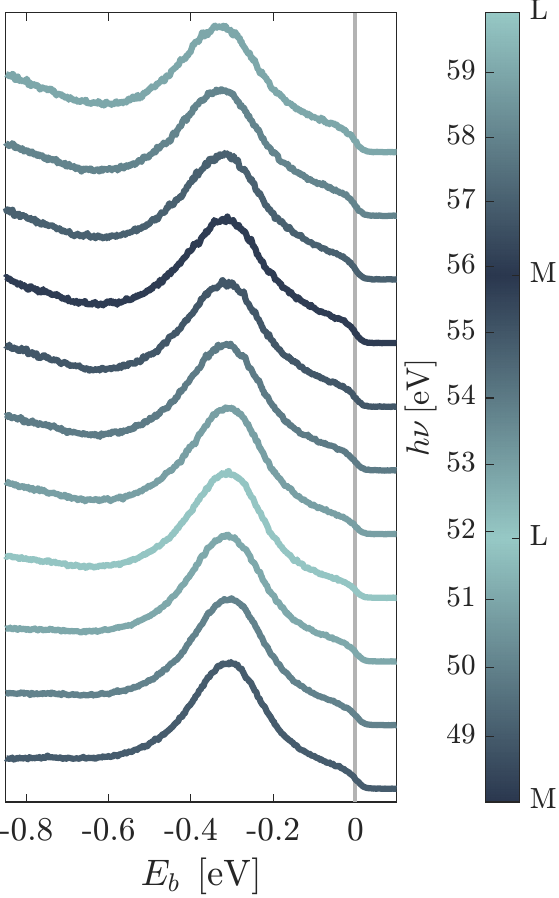}
\caption{Energy distribution curves along the M-L line of the 1H-TaS$_2$ Brillouin zone. The different k$_z$ are measured using different photon energies ranging from \(49\eV\) to \(59\eV\). This range of photon energies covers one and a half BZ along k$_z$. The gray line is a guide to the eye along the Fermi level.}
\label{fig:kzDep}
\end{figure}

By varying the photon energy in an ARPES experiment, we can capture the band structure dispersion with k$_z$~\cite{AlmoalemSpin,Almoalem2021link,Bawden2016}. The lack of noticeable change of the bands along the M-L line indicates an absence of a gap between the bands at the Brillouin zone M-point, Supp. Fig. \ref{fig:kzDep}. A maximum gap is predicted to exist in the M-point due to an inter layer coupling term in the tight-binding model~\cite{AlmoalemSpin}. Therefore, the uniform structure of the bands points to a vanishing inter layer coupling between the 1H layers, effectively creating separated 1H monolayers in the sample. 

\newpage
\section{Theoretical model}

In this supplement we analyze the effect of interlayer coupling on the triplet-singlet mixing in TMD superconductors of the H type (2H-TaS$_2$, 2H-NbSe$_2$, 2H-TaSe$_2$, etc.). We will consider a  single bilayer with finite-ranged attractive interactions and tune the strength of the interlayer hopping.  We then investigate how the ``mutli-gap'' features, e.g. in the tunneling density of states, depend on the coupling. 

\subsection{Model}
For simplicity we consider a toy model that contains much of the physics of a single bilayer of 2H-TaS$_2$ with a finite range intralayer  attractive interaction 
\begin{align}\label{eq:H}
H = \sum_{\bs k} c_{\bs k}^\dagger \left[ t\epsilon_{\bs k}-\lambda\, o_{\bs k} \sigma^z \tau^z  - t_\perp \tau_x\right]c_{\bs k} - {1\over 2}\sum_{\bs k \bs p}\sum_{\sigma \sigma'}\sum_{\alpha}V_{\bs k \bs p} c_{\bs k \sigma \alpha}^\dagger c_{-\bs k\sigma' \alpha}^\dagger c_{-\bs p \sigma'\alpha} c_{\bs p \sigma \alpha}
\end{align}
where $\alpha = 1,2$ denotes the layer, $\sigma = \uparrow \downarrow$, denotes spin and the Pauli matrices $\tau^{x,y,z}$ and $\sigma^{x,y,z}$ are in the respective basis. The dispersion functions $\e_{\bs k}$ and $o_{\bs k}$ are given by 
\[\e_{\bs k} = \cos(\bs{T}_1-\bs{T}_2)\cdot \bs k+\cos(\bs{T}_3-\bs{T}_2)\cdot \bs k+\cos(\bs{T}_3-\bs{T}_1)\cdot \bs k\]
\[o_{\bs k} = \sin\bs T_1\cdot \bs k+\sin\bs T_2\cdot \bs k+\sin\bs T_3\cdot \bs k\]
where $\bs T_1 = -a\hat{\bs y}$, $\bs T_2 = a(-\sqrt 3\hat{\bs x}+\hat{\bs y})/2$ and $\bs T_3 = a(\sqrt 3\hat{\bs x}+\hat{\bs y})/2$ are the vectors to the three nearest neighbors. Therefore, $t$ is an energy scale representing the hopping, $\lambda$ is the strength of Ising spin-orbit coupling and $t_\perp$ is the interlayer hopping strength. 
The dispersion part of the Hamiltonian Eq.~\eqref{eq:H} is tailored to have a band structure very similar to the conduction band of TMD metals (see Supplementary Figures 6,7,8). 

The interaction term is assumed to be 
\begin{align} V_{\bs k \bs p} = V_0 + V_1 e_{\bs k - \bs p}+ V_2 \epsilon_{\bs k - \bs p} \end{align}
where 
\[e_{\bs k} = \cos\bs T_1\cdot \bs k+\cos\bs T_2\cdot \bs k+\cos\bs T_3\cdot \bs k\,.\]
Thus, the term $V_0$ represents the on-site interaction, $V_1$ the nearest neighbor interaction and $V_2$ the next-to-nearest neighbor interaction. In what follows, we will see that a finite range interaction is an essential ingredient to get multi-gap features.  

\subsection{BCS theory}
In the spirit of BCS theory we first transform the Hamiltonian to the band basis. This is done using the transformation
\[ c_{\bs k \sigma \alpha} = \sum_{j} U_{\alpha j}{}(\sigma \boldsymbol k)\psi_{\bs k\sigma j} \]
where $\sigma = +1$ for $\uparrow$ and $-1$ for $\downarrow$, $j = \pm$ denotes the bands and
\[ \hat U(\bs k) = \begin{pmatrix} 
\cos{\t_{\bs k}\over2}&\sin{\t_{\bs k} \over 2} \\
\sin{\t_{\bs k}\over 2} & -\cos{\t_{\bs k}\over 2}
\end{pmatrix}\]
Here 
\[ \cos \t_{\bs k} = -{t_\perp \over A_{\bs k}}\;\;;\;\; \sin \t_{\bs k} = -{ \lambda o_{\bs k} \over A_{\bs k}}\;\;
;\;\; A_{\bs k} = \sqrt{\lambda^2 o_{\bs k}^2 + t_\perp^2}\]
Using this transformation we find that the Hamiltonian Eq.~\eqref{eq:H} assumes the form 
\begin{align}\label{eq:H2}
H = \sum_{\bs k\sigma j} \zeta_{\bs k j}\,\psi_{\bs k\sigma j}^\dagger \psi_{\bs k\sigma j} - \sum_{\bs k \bs p}\sum_{\sigma \sigma'}\sum_{j_1 j_2 j_3 j_3}\tilde V_{\bs k \bs p;\s \s'}^{j_1 j_2 j_3 j_4} \psi_{\bs k \sigma j_1}^\dagger \psi_{-\bs k\sigma' j_2}^\dagger \psi_{-\bs p \sigma'j_3} \psi_{\bs p \sigma j_4}
\end{align}
where $\zeta_{\bs k \pm} = t\e_{\bs k}\pm A_{\bs k}$ and
\[\tilde V_{\bs k \bs p;\s\s'}^{j_1 j_2 j_3 j_4} ={1\over 2} \sum_\alpha U_{\alpha j_1}^*(\sigma \bs k)U_{\alpha j_2}^*(-\sigma' \bs k) U_{\alpha j_3}(-\sigma' \bs p)U_{\alpha j_4}(\sigma \bs p)V_{\bs k \bs p}\]
Here we will limit ourselves to \emph{intraband pairing only}. Furthermore, due to the Ising spin-orbit coupling we consider only pairing states between spin up at one momentum and spin down in the opposite momentum (spin along $z$ remains a good quantum number). Consequently, the following product 
\[ U_{\alpha j}(-\sigma' \bs k)U_{\alpha \ell}(\sigma \bs k) = U_{\alpha j}( \bs k)U_{\alpha \ell}( \bs k) \]
becomes spin-independent and so does the interaction. 
We then decompose the interaction into symmetric and anti-symmetric parts
\[\tilde V_{\bs k \bs p;j\ell}^{(s)} ={1\over 4} \left[ \tilde V_{\bs k \bs p}^{jj\ell\ell}+\tilde V_{-\bs k \bs p}^{jj\ell\ell}+\tilde V_{\bs k ,-\bs p}^{jj\ell\ell}+\tilde V_{-\bs k ,-\bs p}^{jj\ell\ell}
\right]\]
\[\tilde V_{\bs k \bs p;j\ell}^{(a)} ={1\over 4} \left[ \tilde V_{\bs k \bs p}^{jj\ell\ell}-\tilde V_{-\bs k \bs p}^{jj\ell\ell}-\tilde V_{\bs k ,-\bs p}^{jj\ell\ell}+\tilde V_{-\bs k ,-\bs p}^{jj\ell\ell}
\right]\]
and use the following BCS mean-field decoupling to the interaction term,
\begin{align}\label{eq:Self_consistent}
\begin{split}
\Delta^{(s)}_j (\bs k)&= \sum_{\bs p \ell} V_{\bs k \bs p;j\ell}^{(s)} \left[\langle \psi_{-\bs p \downarrow \ell}\,\psi_{\bs p \uparrow \ell} \rangle - \langle \psi_{-\bs p\uparrow \ell}\,\psi_{\bs p\downarrow \ell} \rangle \right]\\
\Delta^{(a)}_j (\bs k)&= \sum_{\bs p \ell} V_{\bs k \bs p;j\ell}^{(s)} \left[\langle \psi_{-\bs p \downarrow \ell}\,\psi_{\bs p \uparrow \ell} \rangle + \langle \psi_{-\bs p\uparrow \ell}\,\psi_{\bs p\downarrow \ell} \rangle \right]\,.
\end{split}
\end{align}
The above decoupling leads to the following BdG Hamiltonian 
\begin{align}\label{eq:H_BdG}
H_{\text{BdG}}^{j} = \begin{pmatrix} \zeta_{\bs kj} & -\D_j^{(s)}(\bs k)-\D_{j}^{(a)}(\bs k) & & \\
-\D_j^{(s)}(\bs k)-\D_{j}^{(a)}(\bs k) & -\zeta_{-\bs kj} & & \\
& & \zeta_{\bs kj} & -\D_j^{(s)}(\bs k)+\D_{j}^{(a)}(\bs k) \\ & & -\D_j^{(s)}(\bs k)+\D_{j}^{(a)}(\bs k)& -\zeta_{-\bs kj}
\end{pmatrix}
\end{align}
which is written in the Nambu basis $\Psi_{\bs k} = [\psi_{\bs k \uparrow j},\psi_{-\bs k \downarrow j}^\dagger,\psi_{\bs k \downarrow j},-\psi_{-\bs k \uparrow j}^\dagger ]^T$.  Using the above Hamiltonian we can compute the expectation values for the self-consistency equations 
\begin{align}
\langle \psi_{-\bs p \downarrow j}\,\psi_{\bs p \uparrow j} \rangle = -{\Delta_j^{(s)}(\bs k)+\Delta_j^{(a)}(\bs k) \over \sqrt{\zeta_{\bs k j}^2+\left[\Delta_j^{(s)}(\bs k)+\Delta_j^{(a)}(\bs k)\right]^2}}\\
\langle \psi_{-\bs p \uparrow j}\,\psi_{\bs p \downarrow j} \rangle = -{\Delta_j^{(s)}(\bs k)-\Delta_j^{(a)}(\bs k) \over \sqrt{\zeta_{\bs k j}^2+\left[\Delta_j^{(s)}(\bs k)-\Delta_j^{(a)}(\bs k)\right]^2}}
\end{align}

\subsection{Results}
In what follows we describe the results of our simulation. We first review the properties of the normal state. We then turn to the gap structure and finally use it to compute the tunneling DOS. 

We show that under the right circumstances the isolated monolayer ($t_\perp = 0$) can develop huge anisotropy in momentum space, resulting from singlet-triplet mixing, which in turn, leads to strong ``multi-gap'' features in the tunneling density of states. The anisotropy and singlet-triplet mixing diminish  upon increasing the interlayer coupling $t_\perp$, which is consistent with the body of experimental work. 

Two ingredients are essential for this behavior to appear: (i) The attractive interaction needs to be of finite range. In particular, the on-site attraction needs to be comparable, or smaller than, the nearest neighbor and next-to-nearest neighbor interaction. This may occur due to repulsive Coulomb interaction, which tends to be shorter ranged than the phonon attraction. (ii) Breaking of inversion symmetry, which in this case is manifested by the Ising spin-orbit coupling. Without, the breaking of inversion the singlet-triplet mixing is prohibited and the effect disappears. 

\subsubsection{Dispersion}

\begin{figure*}[!h]\label{fig:disp_t=0}
\includegraphics[trim= 0cm 0cm 0cm 0cm,clip=true,width=1\textwidth]{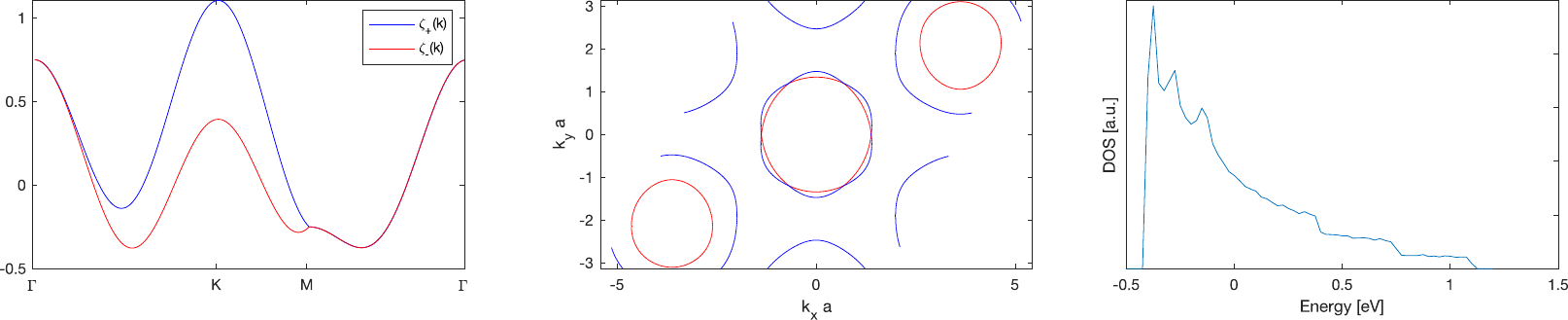}
\caption{Left: The dispersion functions of the model Hamiltonian Eq.~\eqref{eq:H2}, $\z_\pm$ as a function of momentum for the case of $t=0.25$ eV, $\lambda = 0.3$ eV and $t_\perp = 0$. Middle: The corresponding Fermi surfaces in the 1$^{\text{st}}$BZ. Right: The single particle density of states as a function of energy.}
\end{figure*}

\begin{figure*}[!h]
\includegraphics[trim= 0cm 0cm 0cm 0cm,clip=true,width=1\textwidth]{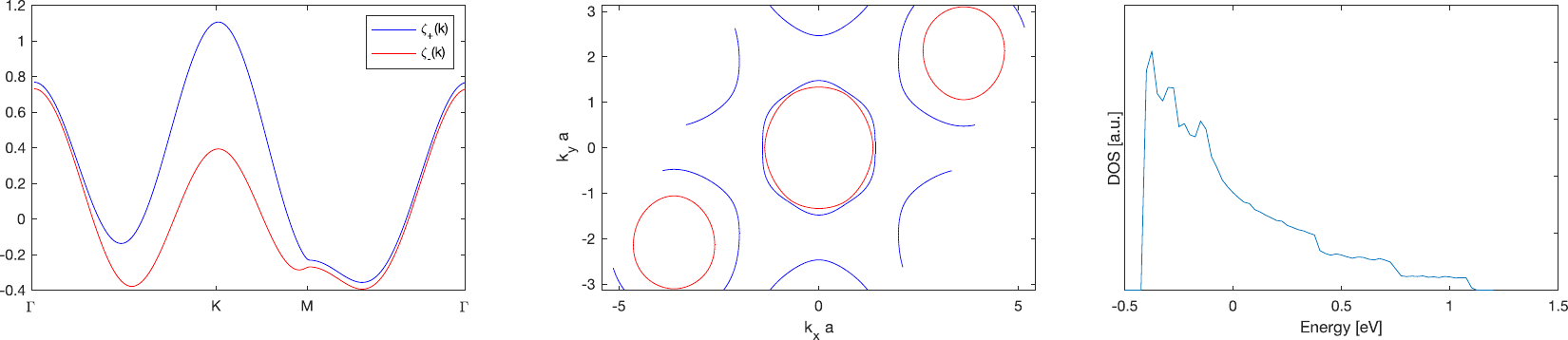}
\caption{Left: The dispersion functions of the model Hamiltonian Eq.~\eqref{eq:H2}, $\z_\pm$ as a function of momentum for the case of $t=0.25$ eV, $\lambda = 0.3$ eV and $t_\perp = 0.25 \lambda$. Middle: The corresponding Fermi surfaces in the 1$^{\text{st}}$BZ. Right: The single particle density of states as a function of energy.}
\label{fig:disp_t=0.25}
\end{figure*}

\begin{figure*}[!h]
\includegraphics[trim= 0cm 0cm 0cm 0cm,clip=true,width=1\textwidth]{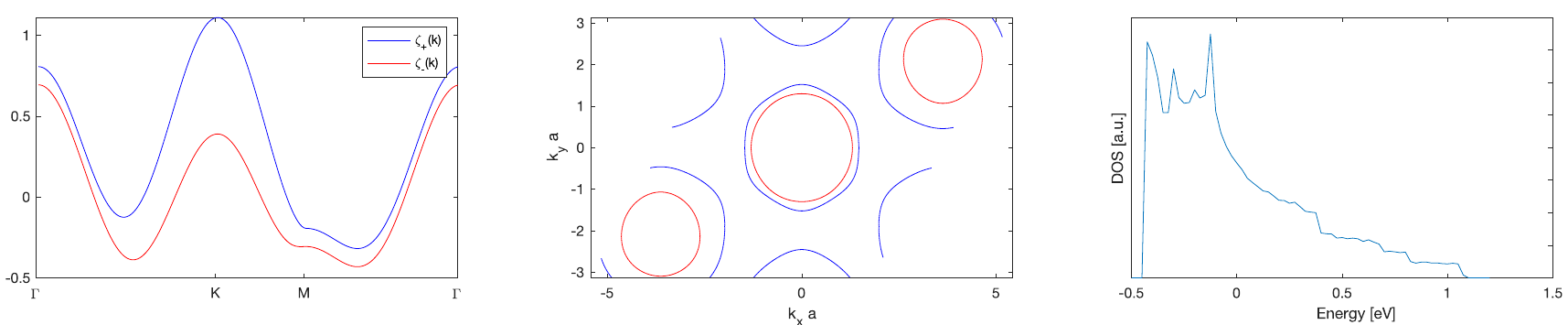}
\caption{ Left: The dispersion functions of the model Hamiltonian Eq.~\eqref{eq:H2}, $\z_\pm$ as a function of momentum for the case of $t=0.25$ eV, $\lambda = 0.3$ eV and $t_\perp = 0.75 \lambda$. Middle: The corresponding Fermi surfaces in the 1$^{\text{st}}$BZ. Right: The single particle density of states as a function of energy.}
\label{fig:disp_t=0.75}
\end{figure*}

In Figs.~\ref{fig:disp_t=0}, \ref{fig:disp_t=0.25},\ref{fig:disp_t=0.75} we plot the dispersion functions of the model Hamiltonian Eq.~\eqref{eq:H2}, $\z_\pm$ as a function of momentum for the case of $t=0.25$ eV, $\lambda = 0.3$ eV and $t_\perp = 0,\,0.25\lambda\,\&\,0.75\lambda$ on the left. The corresponding Fermi surfaces in the 1$^{\text{st}}$BZ in the middle panels and the single particle density of states as a function of energy in the right panels.
As can be seen, the model qualitatively mimcs the features of the band structure in H structured TMD metals. 

\subsubsection{Gap structure and Tunneling density of states}
\begin{figure*}[!h]
\includegraphics[width=1\textwidth]{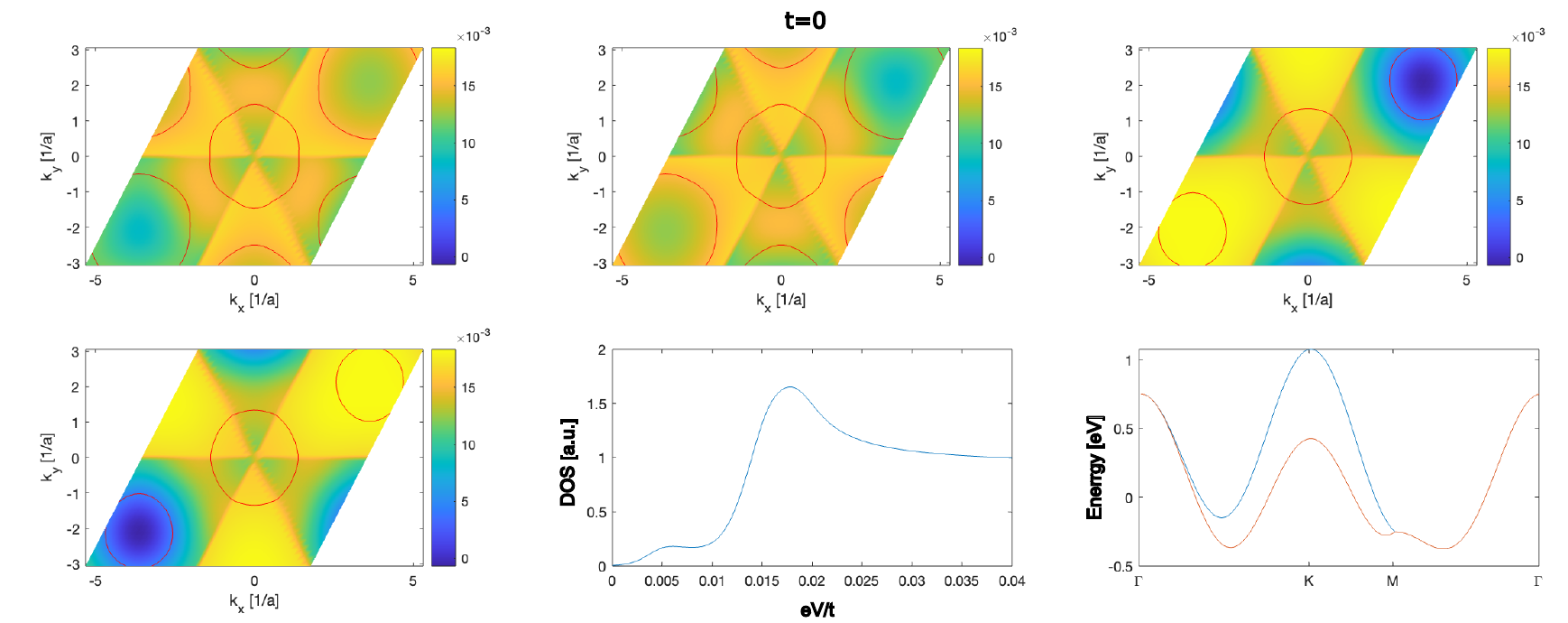}
\caption{The three top panels and the lower left panel are a plot of the gap function in the first Brillouin zone for $t_\perp = 0$, $\lambda = 0.3$ eV, $t=0.25 $ eV, $V_0 = 0.25 t$, $V_1 = V_2 = 0.3t$. From top left, $\Delta_+^{(s)}+\Delta_+^{(a)}$, $\Delta_+^{(s)}-\Delta_+^{(a)}$, $\Delta_-^{(s)}+\Delta_-^{(a)}$ and $\Delta_-^{(s)}-\Delta_-^{(a)}$, respectively. The remaining two plots are the corresponding tunneling density of states and band-structure.}
\label{fig:gap_t=0}
\end{figure*}

\begin{figure*}[!h]
\includegraphics[trim= 0cm 0cm 0cm 0cm,clip=true,width=1\textwidth]{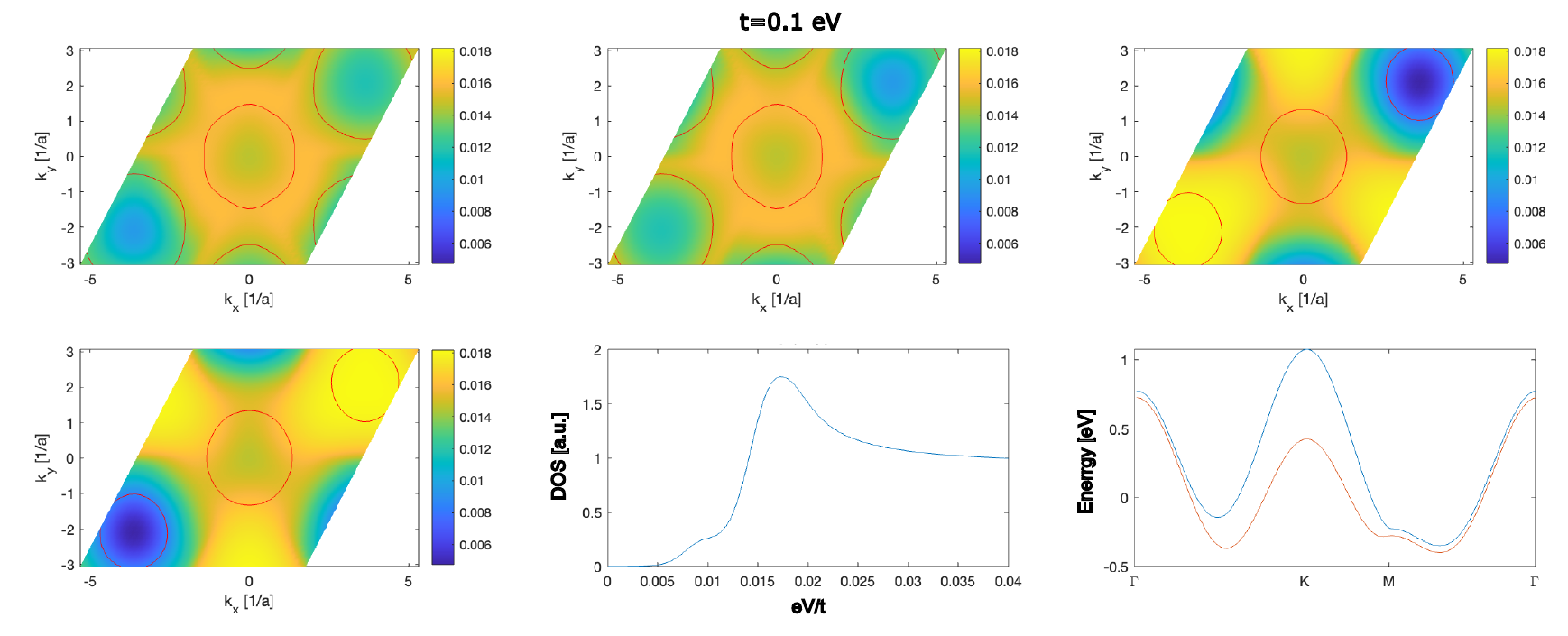}
\caption{The three top panels and the lower left panel are a plot of the gap function in the first Brillouin zone for $t_\perp = 0.1$ eV, $\lambda = 0.3$ eV, $t=0.25 $ eV, $V_0 = 0.25 t$, $V_1 = V_2 = 0.3t$. From top left, $\Delta_+^{(s)}+\Delta_+^{(a)}$, $\Delta_+^{(s)}-\Delta_+^{(a)}$, $\Delta_-^{(s)}+\Delta_-^{(a)}$ and $\Delta_-^{(s)}-\Delta_-^{(a)}$, respectively. The remaining two plots are the corresponding tunneling density of states and band-structure. }
\label{fig:gap_t=0.1}
\end{figure*}

\begin{figure*}[!h]
\includegraphics[trim= 0cm 0cm 0cm 0cm,clip=true,width=1\textwidth]{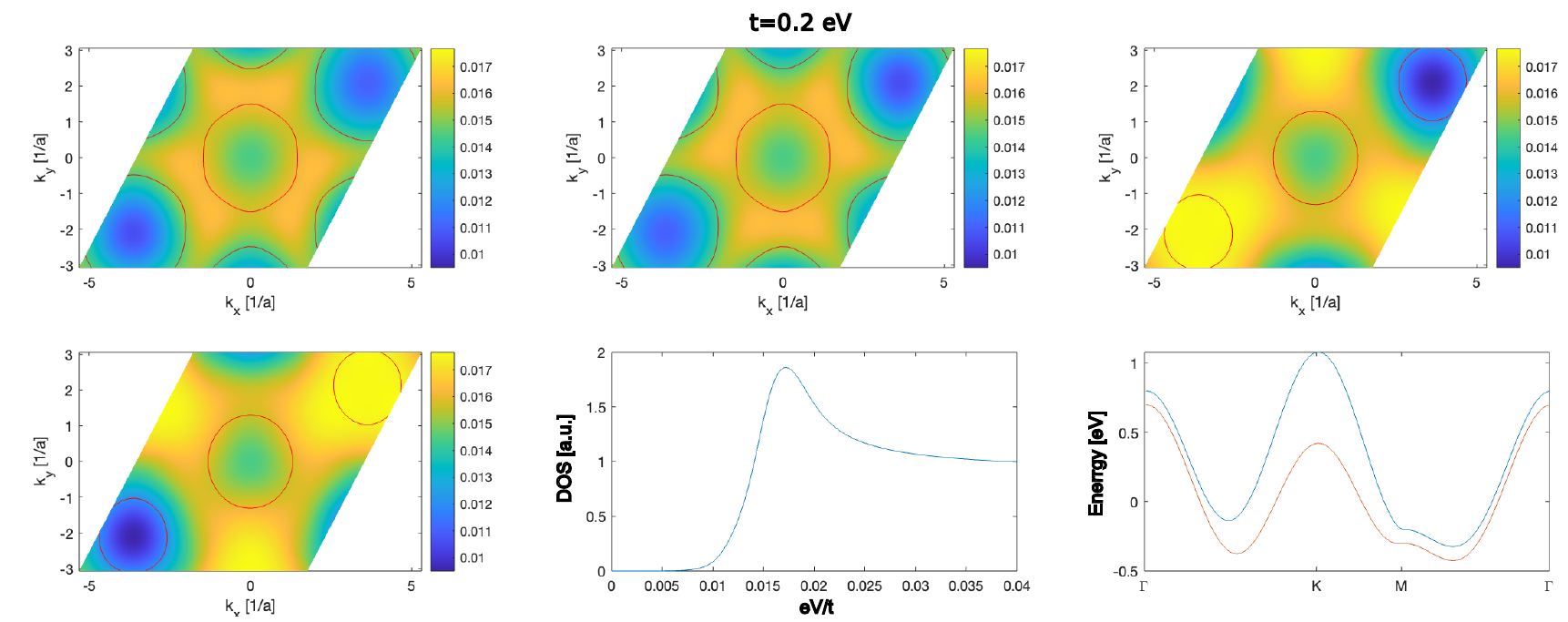}
\caption{The three top panels and the lower left panel are a plot of the gap function in the first Brillouin zone for $t_\perp = 0.2$ eV, $\lambda = 0.3$ eV, $t=0.25 $ eV, $V_0 = 0.25 t$, $V_1 = V_2 = 0.3t$. From top left, $\Delta_+^{(s)}+\Delta_+^{(a)}$, $\Delta_+^{(s)}-\Delta_+^{(a)}$, $\Delta_-^{(s)}+\Delta_-^{(a)}$ and $\Delta_-^{(s)}-\Delta_-^{(a)}$, respectively. The remaining two plots are the corresponding tunneling density of states and band-structure.}
\label{fig:gap_t=0.2}
\end{figure*}

In Figs.~\ref{fig:gap_t=0}, \ref{fig:gap_t=0.1} and \ref{fig:gap_t=0.2} we plot the results of a self-consistent calculation of Eq.~\eqref{eq:Self_consistent} over the entire Brillouin zone. 
All figures contain the four gap functions in the Hamiltonian Eq.~\eqref{eq:H_BdG}. Then the last two panels are the tunneling density of states and the corresponding band structure. 

For all three cases we use a finite range interaction $V_0 = 0.25 t$ and $V_1 = V_2 = 0.3t$, where we assume the short range attraction is diminished due to Coulomb repulsion. 
For the momentum space integrals we use a Monkhorst-Pack Grid with 40 points in each direction. 
For the tunneling density of states calculation we use an Allen-Dynes broadening parameter $\Gamma = 1$ meV. 

Fig.~\ref{fig:gap_t=0} shows the case of two isolated H layers $t_\perp = 0$. Here we find a large singlet-triplet mixing. The ratio between the small and big gap on the small $K$ and $K'$ pockets is of the order 0.1. Consequently, we clearly see two gaps in the tunneling density of states.  

Fig.~\ref{fig:gap_t=0.1} shows the case of two isolated H layers $t_\perp = 0.1$ eV. We observe that the increase of the singlet-triplet mixing has diminished due to the increasing $t_\perp$, as reflected by the by the weaker anisotropy in the sizes of the gaps across the BZ. 

Finally, in Fig.~\ref{fig:gap_t=0.2} we show the case of two isolated H layers $t_\perp = 0.2$ eV. We observe that the second gap feature in the tunneling density of states has disappeared due to the decreasing singlet-triplet mixing.

\subsubsection{Conclusions}

We have demonstrated how a multi-gap feature can appear in the Bogoliubov spectrum due to singlet-triplet mixing in a single layer of 2H-TaS$_2$. Furthermore, we have shown how the interlayer coupling between the two constituate layers suppresses the mixing and by that the multi-gap spectral signiture. 

Here it is important to make a comment. First, in principle, a superposition of the gap functions $1$ and $e(\boldsymbol{k})$ can also lead to a variation of the gap (between the $K/K'$ pockets and the $\Gamma$ pocket). However, unlike in the singlet-triplet mixing scenario, this type of mixing is not expected to depend strongly on interlayer coupling. 

\newpage
\bibliographystyle{apsrev4-2}
\bibliography{MainBib}